\documentclass[a4paper,11pt]{article}

\usepackage[dvips]{graphicx}
\usepackage{amsmath,amssymb,float}

\newcommand{\be}{\begin{equation}}
\newcommand{\ee}{\end{equation}}
\newcommand{\ba}{\begin{array}{c}}
\newcommand{\ea}{\end{array}}
\newcommand{\bqaa}{\begin{eqnarray*}}
\newcommand{\eqaa}{\end{eqnarray*}}
\newcommand{\cO}{{\cal O}}
\newcommand{\mL}{\mathcal{L}}

\newcommand{\bqa}{\begin{eqnarray}}
\newcommand{\eqa}{\end{eqnarray}}

\begin{document}
\title{\bf Study of $\eta-\eta'$ mixing from  radiative decay processes}

\author{Yun-Hua Chen$^\dagger$, 
Zhi-Hui~Guo$^{\ddagger,\S,}$\thanks{guo@um.es}, 
Han-Qing Zheng$^\dagger$ 
\vspace{0.3cm} \\
{\small $^\dagger$  Department of Physics and State Key Laboratory of Nuclear Physics and Technology, } \\
{\small Peking University, Beijing 100871, P.~R.~China.  }
\\ {\small $^\ddagger$ Department of Physics, Hebei Normal University, 050016 Shijiazhuang, P.~R.~China.}
\\ {\small $^\S$ Departamento de F\'{\i}sica, Universidad de Murcia, E-30071 Murcia, Spain.}  }

\date{}
\maketitle

\begin{abstract}
We perform a thorough analysis of the $VP\gamma(\gamma^*)$ and
$P\gamma\gamma(\gamma^*)$ decays in the resonance chiral theory, 
where $V$ stand for the vector resonances $\rho, K^*, \omega,
\phi$, $P$ stand for $\pi, K, \eta, \eta'$ and $\gamma^*$
subsequently decays into lepton pairs. Upon imposing QCD
short-distance constraints on resonance couplings, the $\omega \to
\pi \gamma(\gamma^*)$, $\rho \to \pi \gamma(\gamma^*)$, $K^{*0} \to
K^0 \gamma$ processes only depend on one free parameter and 
$\pi \to \gamma \gamma(\gamma^*)$ can be completely predicted. 
The four mixing parameters of the 
$\eta-\eta'$ system, i.e. two mixing angles $\theta_8, \theta_0$ and
two decay constants $F_8, F_0$, are determined from radiative decays
involving  $\eta$ or $\eta'$. The higher order low energy constants
of the pseudo-Goldstone Lagrangian in the chiral anomaly sector are
predicted by integrating out heavy resonances. We also
predict the decay widths of ${\rho\rightarrow \pi e^{+}e^{-}}$,
${\eta'\rightarrow \gamma e^{+}e^{-}}$ and ${\phi\rightarrow \eta
\mu^{+}\mu^{-}}$, which can be compared with the  future measurement in
these channels.
\end{abstract}


\section{Introduction}

To study the properties of $\eta$ and $\eta'$ mesons is a very
interesting subject in hadron physics. The reasons behind are
twofold. First, several experimental collaborations have started or
planned programs to launch the measurements of the processes
involving the $\eta$ and $\eta'$ mesons with high statistics and
high precision, such as KLOE~\cite{kloe}, Jefferson
Lab~\cite{jlab} and BES-III~\cite{besiii}. The huge data sample, for
example 63 million events for $\eta$ decays and 61 million events
for $\eta'$ decays expected at BES-III~\cite{besiii}, apparently
needs more and finer theoretical work for the analysis. Second, on
the theoretical side, $\eta$ and $\eta'$ mesons present important
information of low energy dynamics of QCD: the mechanism of
spontaneously chiral symmetry breaking and the $U_A(1)$ anomaly.

The responsible theory in the low energy region of QCD is Chiral
Perturbation Theory ($\chi$PT)~\cite{gasser8485}, whose degrees of
freedom are the pseudo-Goldstone mesons, i.e. $\pi, K, \eta$,
resulted from the spontaneously chiral symmetry breaking from
${SU(3)}_L \bigotimes {SU(3)}_R$ to $SU(3)_{V=L+R}$. $\chi$PT has
been proved to be a very successful effective field theory to
describe the low energy physics of QCD~\cite{chptreview}, that is
constructed with respect to chiral symmetry and arranges its
effective action in the expansion of momenta and the mass of
pseudo-Goldstone bosons. Due to  $U_A(1)$ anomaly, $\eta'$ is
prevented to be the ninth pseudo-Goldstone boson. Nevertheless, when
the number of the colors in QCD, $N_C$, becomes large, the effect of
$U_A(1)$ anomaly is  suppressed. Thus the $\eta'$ meson becomes the
ninth pseudo-Goldstone boson in the large $N_C$ limit~\cite{ua1innc}
and can be incorporated into a chiral Lagrangian. Such an effective
theory explicitly including the $\eta'$ meson extends the standard
${SU(3)}_L \bigotimes {SU(3)}_R$ $\chi$PT to the ${U(3)}_L
\bigotimes {U(3)}_R$ version, whose Lagrangian up to $O(p^4)$ has
been thoroughly investigated in~\cite{Kaiser00,Siklody} for the
even intrinsic parity sector.

The application of the ${U(3)}_L \bigotimes {U(3)}_R$ $\chi$PT to
$\eta-\eta'$ mixing has been performed in the
literature~\cite{Siklody98,Leutwyler98,Kaiser98}, where the mixing
parameters of $\eta$ and $\eta'$ are determined by including higher
order corrections, such as the low energy constants(LECs) and the
loops. A novel finding after including the higher order contributions
in $\chi$PT is that the conventional one-mixing-angle description for
$\eta-\eta'$ mixing is not valid any more and two-mixing-angle scheme is
then proposed in~\cite{Leutwyler98}.
In our current discussion, instead of working in more detail
in the top-down method to address the $\eta-\eta'$ mixing problem,
we are going to determine the mixing couplings in a bottom-up way,
i.e., we assume the validity of the two-mixing-angle description and
then directly fit them using experimental data from the relevant
physical processes that involve $\eta$ or $\eta'$. We will focus on
the radiative decay processes in the present work, since they can
provide a large sample of data~\cite{pdg} that allows us to better
extract the $\eta-\eta'$ mixing information and they are less
contaminated by the strong final state interaction comparing with
the hadronic decay modes, such as $\eta' \to \eta\pi\pi$
~\cite{escribano2011jhep}.

Among the radiative decay processes with $\eta$ or $\eta'$ meson in
the low energy sector, many of them consist of one vector resonance,
such as $\phi \to \eta \gamma$, $\eta' \to \omega \gamma$, and so
on~\cite{pdg}. Apparently, these  processes are already beyond the
validity region of $\chi$PT due to the appearance of the heavy
vector resonances. It is by no means trivial to systematically
include the heavy vector resonances in $\chi$PT, since the expansion
parameters of $\chi$PT are no longer valid after the inclusion of
the heavy multiplet of resonances. Nevertheless, the framework developed in
Ref.~\cite{Ecker}, named Resonance Chiral Theory (R$\chi$T), has
been proven to be useful and may shed light on the proper
construction of the Lagrangian theory that one could use to describe
the dynamics with both pseudo-Goldstone mesons and resonances. It
can be better understood within the framework of the large $N_C$ QCD
as theory of hadrons~\cite{PerisdRafael}. While in the strict large
$N_C$ QCD there is an infinite number of zero-width hadrons in the
spectrum, in practical realization one usually needs to truncate the
infinite tower of resonances to the lowest multiplet for each
quantum number. There has been a large amount of research works
based on this approximation, varying from determination of the
$\chi$PT LECs~\cite{kampf-06,guo-09} to the study of tau
decays~\cite{guo-08,gomez-dumm-10,guo-10} and Green functions of QCD
currents~\cite{gf-moussallam-95,gf-moussallam-97,gf-bijnens,gf-knecht,
gf-pich-03}.

R$\chi$T, although well respecting  chiral symmetry and constructed in the
guide of $1/N_C$ expansion, is still lacking of QCD dynamics at the high energy
scale where the continuum is reached and perturbative QCD is the
responsible theory. Thus it is crucial to match the behaviors of the
effective field theory and QCD at high momentum transfer to
implement as many QCD features as possible. Research along this line
indeed has been intensively performed in many works
\cite{gf-moussallam-95,gf-moussallam-97,gf-bijnens,gf-knecht,
gf-pich-03}. This procedure directly results in the constraints on
the resonance couplings and hence makes the theory more predictable.

In the present work, we utilize  R$\chi$T to  analyze  the
$VP\gamma(\gamma^*)$ and $P\gamma\gamma(\gamma^*)$ processes.
Comparing with the work in Ref.~\cite{gf-pich-03}, we generalize the
R$\chi$T Lagrangian with the octet of pseudo-Goldstone mesons to the
version with nonet, thus allowing us to study the processes with
$\eta'$ meson. Our work is also especially devoted to the
determination of the $\eta-\eta'$ mixing parameters by fitting
data, which we will explain in detail later in the text.

We organize the article as follow. A mini-review on $\eta-\eta'$
mixing is given in Section~\ref{background-eta}. The structure of
the relevant R$\chi$T Lagrangian is elaborated in
Section~\ref{rxt-intro}. The computation of the decay amplitudes of
$VP\gamma(\gamma^*)$ and $P\gamma\gamma(\gamma^*)$ is noted in
Section~\ref{calc-amp}. The QCD short distance constraints are
discussed in Section~\ref{he-constraint}. Phenomenology discussion
is given in Section~\ref{pheno} and we conclude in
Section~\ref{conclude}.

\section{ Mini-review on $\eta-\eta'$ mixing}\label{background-eta}

$\eta-\eta'$ mixing is  an interesting  subject in hadron physics.
In the literature, the mixing angles have been defined with respect
to different bases: the octet-singlet flavour basis and the quark
flavour basis. For our purpose, we will always adopt the
octet-singlet flavour basis to define the mixing angles throughout
this article. The two-mixing-angle description has been proposed to
settle the $\eta-\eta'$ mixing~\cite{Leutwyler98}, going beyond the
old one-mixing-angle description~\cite{Feldmann00}. The requirement
of the two mixing angles can be better understood in the $\chi$PT
frame. The leading order Lagrangian of ${U(3)}_L \bigotimes
{U(3)}_R$ $\chi$PT is
\bqa \label{loLagrangian} \mL^{(0)}=\frac{
F^2}{4}\langle u_\mu u^\mu \rangle+ \frac{F^2}{4}\langle \chi_+
\rangle + \frac{F^2}{3}M_0^2 \ln^2{\det u}\,, \eqa where
\be\label{defu}
       u(x)= \exp(i\frac{ \Phi(x)}{ \sqrt2F})\,,
\ee
\bqa
\label{phi1}
\Phi(x) \,=\, \left( \begin{array}{ccc}
\frac{\sqrt 3 \pi^0+\eta_8+\sqrt 2 \eta_1}{\sqrt 6} & \pi^+ & K^+  \\ \pi^- &
\frac{-\sqrt 3 \pi^0+\eta_8+\sqrt 2 \eta_1}{\sqrt 6}   & K^0 \\  K^- & \bar{K}^0 &
\frac{-2\eta_8+\sqrt 2 \eta_1 }{\sqrt 6}
\end{array} \right)\,.
\eqa The last term in Eq.(\ref{loLagrangian}) represents the
$U_A(1)$ anomaly of QCD, which gives rise to the singlet $\eta_1$
mass, $M_0$. $F$ is the value of the pseudo-Goldstone decay constant
in the chiral limit, with the normalization of $F_\pi = 92.4$ MeV.
For other chiral building blocks, see Ref.~\cite{Ecker} and
references therein.

At leading order, only one mixing angle is needed to diagonalize the
octet $\eta_8$ and the singlet $\eta_1$ to get the mass eigenstates
$\eta$ and $\eta'$, since there is only one mixing term in the mass
sector. However, the higher order corrections, including the loop
contributions and the LECs from the higher order Lagrangian,
contribute not only to the mixing of mass but also to the mixing of
kinetic term. In order to get the physical eigenstates of $\eta$ and
$\eta'$, one needs three steps: first to diagonalize the kinetic
term, then to perform the normalization of each field and to
diagonalize the mass term in the end, which indicates four
parameters, i.e. two angles and two normalization constants, are
needed in this procedure. It has been shown in
Ref.~\cite{Leutwyler98,Kaiser98} the mixing can be parameterized as,
\bqa \label{twoanglesmixing}
 \left(
 \begin{array}{c}
 \eta   \\
 \eta' \\
 \end{array}
 \right) = \frac{1}{F}\left(
                                        \begin{array}{cc}
                                          F_8\, \cos{\theta_8}  & -F_0 \,\sin{\theta_0}  \\
                                           F_8\,\sin{\theta_8} & F_0 \,\cos{\theta_0} \\
                                        \end{array}
    \right)
      \left(
       \begin{array}{c}
       \eta_8   \\
       \eta_1  \\
       \end{array}
        \right)\,,
\eqa where $F_8$ and $F_0$ correspond to the weak decay constants of
the axial octet and singlet currents, respectively. By setting
$F_8=F_0=F$ and $\theta_0=\theta_8$ in Eq.(\ref{twoanglesmixing}),
the conventional one-mixing-angle scheme can be recovered.

The works presented in Refs.~\cite{Leutwyler98,Kaiser98,Siklody98}
are devoted to the determination of the mixing parameters $F_8, F_0,
\theta_8, \theta_0$ by including the higher order corrections in
$U(3)$ $\chi$PT. Recently assorted methods along this line have been
done to settle the $\eta-\eta'$ mixing in \cite{gerard09,
mathieu10}.  As already advertised in the Introduction, instead of
considering more of higher order corrections to calculate the mixing
parameters, we will first adopt the two-mixing-angle scheme
described in Eq.(\ref{twoanglesmixing}) for the $\eta-\eta'$ system
and then determine the unknown mixing parameters phenomenologically.
Similar works within this context have been carried out in
Refs.~\cite{Feldmann97,Feldmann98,Feldmann99,Ball,Escribano99,Pennington,Benayoun00,Goity,
Escribano01,Escribano05,pham10}, which confirm the robustness of the
two-mixing-angle description scheme. In the present work we would
like to readdress the similar processes, such as the $VP\gamma$ and
$P\gamma\gamma$, in the framework of R$\chi$T. The advantage of
R$\chi$T is that it preserves the chiral symmetry in the low energy
sector, respects the high energy behavior of QCD and incorporates
all of the symmetry allowed operators in the construction, in
contrast to the previous works where only a single constant term is
introduced to describe the interaction vertex for $VP\gamma$. In
addition, we perform a global fit by including all of the
experimental available processes with the types of $V \to P \gamma$,
$P \to V \gamma$ and $P \to \gamma \gamma$ and also discuss the
processes with an off-shell photon decaying into lepton pairs: $V
\to P l^+ l^-$, $P \to V l^+ l^-$ and $P \to \gamma l^+ l^-$, where
$V$ stand for $\rho, \omega, \phi, K^*$ and $P$ stand for $\pi, K ,
\eta, \eta'$.

\section{ The relevant Lagrangian of R$\chi$T}\label{rxt-intro}

The resonance chiral effective Lagrangian describing
vector-photon-pseudoscalar and vector-vector-pseudoscalar
vertexes with the vector resonance in the antisymmetric tensor filed
formulation has been given in Ref.~\cite{gf-pich-03}, in which only
the octet of the pseudo-Goldstone mesons is included. Using the
constraints derived from the short distance behavior of QCD, the
Lagrangian has been used to predict the decay widths of $\omega \to
\pi\gamma$ and $\pi\to \gamma\gamma$, which are in good agreement
with the experimental data~\cite{pdg}. In order to study the similar
processes with $\eta'$, we need to generalize the existing resonance
Lagrangian with the pseudo-Goldstone octet to the one involving the
singlet state in addition to the octet. Thanks to  large $N_C$ QCD,
this can be simply accomplished by extending the content of the
unitary matrix $u(x)$, defined in Eq.(\ref{defu}), from the octet to
nonet. In addition, new operators may appear too, as the unitary
matrix $u(x)$ is no longer traceless after the inclusion of the
singlet.

We recall the procedure to construct the $U(3)$ $\chi$PT
Lagrangian~\cite{Kaiser00,Siklody} before illustrating how to build
the new operators in the resonance chiral Lagrangian. The key
ingredient introduced to construct the $U(3)$ $\chi$PT operators is
the $1/N_C$ expansion, in addition to the conventional expansion of
momentum and the quark mass, which is usually named as the triple
expansion scheme, i.e. $\delta  \sim p^2 \sim m_q \sim 1/N_C$. It
also turns out to be useful when building resonance Lagrangian with
the singlet $\eta_1$ as an explicit degree of
freedom~\cite{Kaiser05,Cirigliano06}.

As already mentioned previously, in this article we focus on the
radiative decay processes with the types of $VP\gamma$ and
$P\gamma\gamma$, belonging to the odd intrinsic parity process
describing two vector subjects (photon or vector resonance) and one
pseudoscalar. Guided by the triple expansion scheme, the lowest
order Lagrangian in the odd intrinsic parity sector is in fact the
chiral anomaly formulated in the Wess-Zumino-Witten (WZW)
action~\cite{Wess,Witten2}, with the order of $\cO{(p^4, N_C)}$. The
relevant piece in our discussion can be written in the following way
\begin{eqnarray}\label{lagp4wzw}
\mL_{WZW}=-\frac{\sqrt2 N_C}{8\pi^2 F}
\varepsilon_{\mu\nu\rho\sigma}  \langle \Phi \partial^\mu v^\nu
\partial^\rho v^\sigma \rangle\,,
\end{eqnarray}
where to get the photon field one needs to take $v_\mu = -e\, Q \,
A_\mu$  and the electric charge matrix of the light quarks with
three flavours is $Q = {\rm Diag}\{\frac{2}{3},
-\frac{1}{3},-\frac{1}{3} \}$.

The higher order Lagrangians can be categorized into two types: the
higher order chiral anomaly pseudo-Goldstone Lagrangian and the
Lagrangian with vector resonances. It is known that the higher order
operators in the pure Goldstone Lagrangian encode the information of
heavier degrees of freedom that have been integrated out. To avoid
the double counting in R$\chi$T, the LECs of the higher order
Lagrangian in the pseudo-Goldstone sector
 is usually assumed to be completely saturated by the heavy resonance states and thus
the higher order  operators in the pure pseudo-Goldstone sector can
be dismissed, which works at least pretty well up to the $\cO(p^4)$
level in the even intrinsic sector~\cite{Ecker}. It is pointed out
in Ref.~\cite{Ecker89} to fulfil this procedure it is necessary to
use the antisymmetric formalism to describe the vector resonances.
Though analogous analysis has not been carried out in the odd
intrinsic parity sector, the resonance saturation assumption is
utilized to construct the resonance Lagrangian in
Ref.~\cite{gf-pich-03} and we generalize the discussion by including
the singlet pseudo-Goldstone within the triple expansion scheme.

If the operator is written in terms of $\tilde{u}(x)$, $\tilde{u}(x)
\in U(3)$, it obeys the canonical large $N_C$ counting rules: terms
with a single trace are of order $N_C$ while one additional trace
reduces its order by unity of $1/N_C$. The factor of
$\ln(\det\tilde{u})$ also leads to a suppression of
$1/N_C$~\cite{Kaiser00,Siklody,Kaiser05}. The interacting vertex
involving resonances and pseudo-Goldstones has the general structure
at leading order of $N_C$
\be
 {\cal O}_i \, \sim \, \langle \, R_1 R_2 ... R_j \, \chi^{(n)}(\varphi) \, \rangle \, ,
\ee where $\chi^n(\varphi)$ denotes the chiral tensor that only
incorporates  pseudo-Goldstone bosons and the auxiliary fields with
the chiral order $\cO({p^n})$. For the odd intrinsic parity sector,
it has been shown two types of vector resonance operators are
relevant: $\langle V\chi^{(4)}(\varphi) \rangle$ and $\langle
VV\chi^{(2)}(\varphi) \rangle$ in the case of $u(x) \in SU(3)$. When
the singlet pseudo-Goldstone is taken into account, i.e.
$\tilde{u}(x) \in U(3)$, two new operators with the same chiral
counting order within the triple expansion scheme show up: $\langle
V\chi^{(2)}(\varphi) \rangle \ln(\det\tilde{u})$ and $\langle
VV\rangle \ln(\det\tilde{u})$. The complete Lagrangians are found to be
\begin{eqnarray}
\tilde{\mL}_{VJP}=&\frac{\tilde{c}_1}{M_V}&\varepsilon_{\mu\nu\rho\sigma}
\langle \{V^{\mu\nu},\tilde{f}_+^{\rho\alpha}\}\nabla_\alpha
\tilde{u}^\sigma \rangle \nonumber \\
&+&\frac{\tilde{c}_2}{M_V}\varepsilon_{\mu\nu\rho\sigma}\langle
\{V^{\mu\alpha},\tilde{f}_+^{\rho\sigma}\}\nabla_\alpha
\tilde{u}^\nu\rangle \nonumber\\
&+&\frac{i\tilde{c}_3}{M_V}\varepsilon_{\mu\nu\rho\sigma}
\langle\{V^{\mu\nu},\tilde{f}_+^{\rho\sigma}\} \tilde{\chi}_-\rangle \nonumber\\
&+&\frac{i\tilde{c}_4}{M_V}\varepsilon_{\mu\nu\rho\sigma}\langle
V^{\mu\nu}[\tilde{f}_-^{\rho\sigma},\tilde{\chi}_+]\rangle \nonumber\\
&+&\frac{\tilde{c}_5}{M_V}\varepsilon_{\mu\nu\rho\sigma}\langle \{\nabla_\alpha
V^{\mu\nu},\tilde{f}_+^{\rho\alpha}\}\tilde{u}^\sigma\rangle \nonumber\\
&+&\frac{\tilde{c}_6}{M_V}\varepsilon_{\mu\nu\rho\sigma}\langle \{\nabla_\alpha
V^{\mu\alpha},\tilde{f}_+^{\rho\sigma}\}\tilde{u}^\nu\rangle \nonumber\\
&+&\frac{\tilde{c}_7}{M_V}\varepsilon_{\mu\nu\rho\sigma}\langle \{\nabla^\sigma
V^{\mu\nu},\tilde{f}_+^{\rho\alpha}\}\tilde{u}_\alpha\rangle \nonumber\\
&-&i\tilde{c}_8M_V \sqrt{\frac{2}{3}}\varepsilon_{\mu\nu\rho\sigma}\langle
V^{\mu\nu}\tilde{f}_+^{\rho\sigma}\rangle \ln(\det\tilde{u})\,,
\end{eqnarray}
\begin{eqnarray}
\tilde{\mL}_{VVP}=&\tilde{d}_{1}&\varepsilon_{\mu\nu\rho\sigma}\langle \{V^{\mu\nu},V^{\rho\alpha}\}\nabla_\alpha
\tilde{u}^\sigma\rangle \nonumber\\
&+&i\tilde{d}_{2}\varepsilon_{\mu\nu\rho\sigma}\langle \{V^{\mu\nu},V^{\rho\sigma}\}\tilde{\chi}_-\rangle \nonumber\\
&+&\tilde{d}_{3}\varepsilon_{\mu\nu\rho\sigma}\langle \{\nabla_\alpha
V^{\mu\nu},V^{\rho\alpha}\}\tilde{u}^\sigma\rangle \nonumber\\
&+&\tilde{d}_{4}\varepsilon_{\mu\nu\rho\sigma}\langle \{\nabla^\sigma
V^{\mu\nu},V^{\rho\alpha}\}\tilde{u}_\alpha\rangle \nonumber\\
&-&i\tilde{d}_5M_V^2\sqrt{\frac{2}{3}}\varepsilon_{\mu\nu\rho\sigma}\langle
 V^{\mu\nu}V^{\rho\sigma}\rangle \ln(\det\tilde{u}),
\end{eqnarray}
where we introduce tildes to the objects involving the Goldstone
nonet to distinguish the one with octet. Comparing with
Ref.~\cite{gf-pich-03}, the new operators are
\begin{eqnarray}\label{defc8d5}
\tilde{O}_{VJP}^8&=&-i\tilde{c}_8M_V \sqrt{\frac{2}{3}}\varepsilon_{\mu\nu\rho\sigma}
\langle V^{\mu\nu}\tilde{f}_+^{\rho\sigma}\rangle \ln(\det\tilde{u})\,,\nonumber \\
\tilde{O}_{VVP}^5&=&-i\tilde{d}_5M_V^2\sqrt{\frac{2}{3}}\varepsilon_{\mu\nu\rho\sigma}
\langle V^{\mu\nu}V^{\rho\sigma}\rangle \ln(\det\tilde{u})\,.
\end{eqnarray}
We point out the above operators are only complete for the case with one pseudoscalar field.
In the case with more pseudoscalar states, Ref.~\cite{gf-pich-03} has been
generalized in Ref.~\cite{kampf11prd} to include all of the relevant resonance operators
that can contribute to the $\cO({p^6})$ $\chi$PT LECs in the odd intrinsic parity Lagrangian.

The relevant Lagrangian in the even intrinsic parity sector,
describing the vector resonance and photon transition vertex,
is~\cite{Ecker}
\begin{eqnarray}
\mL_{2}^V=\frac{F_V}{2\sqrt{2}} \langle V_{\mu\nu}\tilde{f}_{+}^{\mu\nu} \rangle \,,
\end{eqnarray}
and the kinetic term for the vector resonance in the antisymmetric formulation reads~\cite{Ecker}
\begin{eqnarray}
\mL_{kin}(V)=-\frac{1}{2} \langle \nabla^{\lambda}V_{\lambda\mu}\nabla_{\nu}V^{\nu\mu}-
\frac{M_V^2}{2}V_{\mu\nu}V^{\mu\nu} \rangle\,,
\end{eqnarray}
where the nonet of the vector resonances resemble the flavor
structure of the pseudo-Goldstone mesons
\begin{equation}
V_{\mu\nu}=
 \left( {\begin{array}{*{3}c}
   {\frac{1}{\sqrt{2}}\rho ^0 +\frac{1}{\sqrt{6}}\omega _8+
   \frac{1}{\sqrt{3}}\omega _1 } & {\rho^+ } & {K^{\ast+} }  \\
   {\rho^- } & {-\frac{1}{\sqrt{2}}\rho ^0 +\frac{1}{\sqrt{6}}\omega _8+
   \frac{1}{\sqrt{3}}\omega _1} & {K^{\ast0} }  \\
   { K^{\ast-}} & {\overline{K}^{\ast0} } & {-\frac{2}{\sqrt{6}}\omega_8
   +\frac{1}{\sqrt{3}}\omega _1 }  \\
\end{array}} \right)_{\mu\nu}.
\end{equation}
For the vector resonances $\omega$ and $\phi$, we assume the ideal
mixing throughout this paper:
\begin{eqnarray}
\omega_1&=&\sqrt{\frac{2}{3}}\omega-\sqrt{\frac{1}{3}}\phi ,  \nonumber \\
   \omega_8&=&\sqrt{\frac{2}{3}}\phi+\sqrt{\frac{1}{3}}\omega.
\end{eqnarray}

The relevant R$\chi$T Lagrangian to our discussion can be summarized
as follow \be \mL = \mL_{WZW} + \mL_{kin}^V + \mL_2^V +
\tilde{\mL}_{VJP}+ \tilde{\mL}_{VVP} \,. \ee

\section{ Theoretical calculation of the radiative decay amplitudes }\label{calc-amp}

In accord with the Lorentz symmetry, the general amplitude for the radiative decay $V(q) \to P(p) \gamma^{*}(k)$
can be written as:
\begin{eqnarray}\label{defFF}
i\mathcal{M}_{V\to P\gamma^{*}}= i \,e\, \varepsilon_{\mu\nu\rho\sigma} \epsilon_V^\mu
\epsilon_{\gamma^{*}}^\nu q^\rho k^\sigma F_{V \to P\gamma^{*}}(Q^2)\,,
\end{eqnarray}
where $\epsilon_V$ and $\epsilon_{\gamma^*}$ denote the polarization
vectors of the vector resonance and the off-shell photon
respectively; the transferred momentum square is defined as $Q^2= -
k^2$. In the case of the on-shell photon, one only needs to replace
$\epsilon_{\gamma^*}$ with the on shell polarization vector
$\epsilon_\gamma$ and impose the real photon condition $Q^2 = - k^2
=0$. Thus the decay widths of $V\to P\gamma$ and $P \to V\gamma$ are
found to be:
\begin{eqnarray}
\Gamma(V\rightarrow
P\gamma)=\frac{1}{3}\alpha(\frac{M_V^2-M_P^2}{2M_V})^3|F_{V\rightarrow
P\gamma^{*}}(0)|^2\,,
\end{eqnarray}
\begin{eqnarray}
\Gamma(P\rightarrow
V\gamma)=\alpha(\frac{M_P^2-M_V^2}{2M_V})^3|F_{P\rightarrow
V\gamma^{*}}(0)|^2\,,
\end{eqnarray}
where $\alpha =e^2/4\pi$ stands for the fine structure constant and the form factor
$F_{P\rightarrow V\gamma^{*}}(Q^2)$ can be defined in the same way as Eq.(\ref{defFF})
\begin{eqnarray}\label{defFF2}
i\mathcal{M}_{P(p) \to V(q) \gamma^{*}(k)}= i \,e\, \varepsilon_{\mu\nu\rho\sigma} \epsilon_V^\mu
\epsilon_{\gamma^{*}}^\nu {q}^\rho {k}^\sigma F_{P \to V\gamma^{*}}(Q^2=-k^2)\,.
\end{eqnarray}

The decay width of $P\rightarrow \gamma\gamma$ can be calculated
\begin{eqnarray}\label{defpggwidth}
\Gamma(P\rightarrow \gamma\gamma)=\frac{1}{4}\pi
\alpha^2M_P^3|F_{P\rightarrow \gamma\gamma^{*}}(0)|^2\,,
\end{eqnarray}
where the form factor $F_{P\rightarrow \gamma\gamma^{*}}(Q^2)$ is defined in an analogous way as Eq.(\ref{defFF2}) by
replacing the vector resonance $V$ with an on-shell photon.

\begin{figure}[ht]
\centering
\includegraphics[angle=0,width=1.0\textwidth]{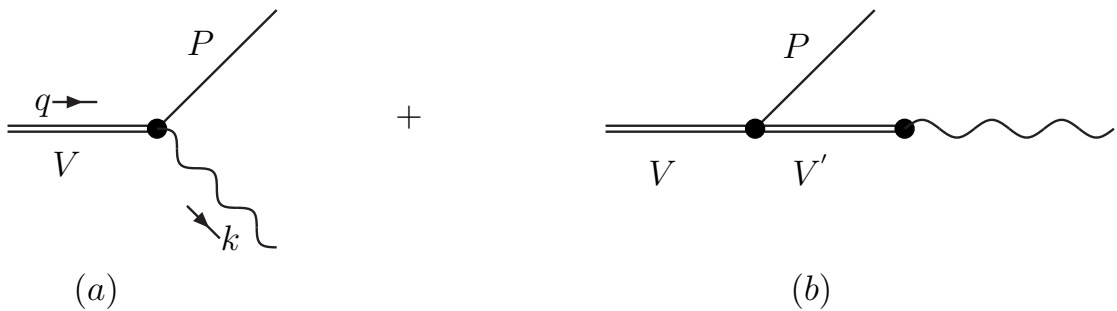}
\caption{ Two types of Feynman diagrams for the processes $VP\gamma^{*}$. }\label{vpgfd}
\end{figure}

\begin{figure}[ht]
\includegraphics[angle=0,width=1.0\textwidth]{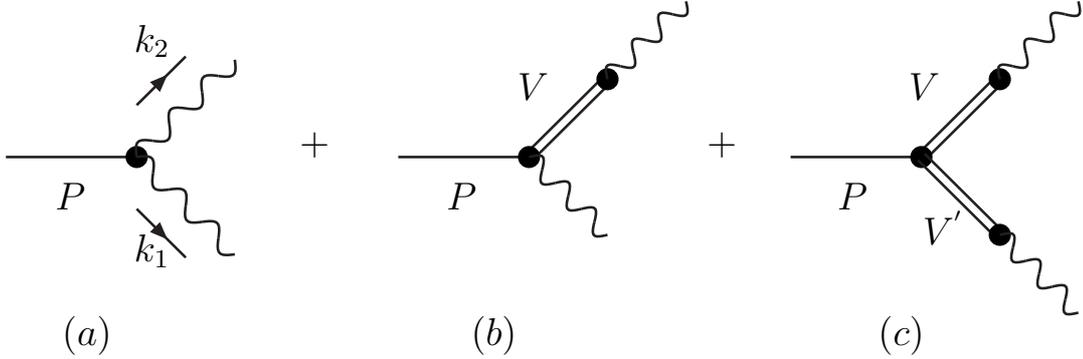}
\caption{ Feynman diagrams for $P\gamma \gamma^*$  }\label{pggfd}
\end{figure}

Next we use the resonance chiral Lagrangian described previously to
calculate the decay widths and transition form factors defined
above, which represent one of our main results in this work. We only
consider the tree-level amplitudes here. The relevant Feynman
diagrams to the radiative $VP\gamma^{*}$ transitions are displayed
in Fig.~\ref{vpgfd}. It is worth pointing out in the celebrated
vector meson dominant(VMD) model only the type (b) diagram in
Fig.~\ref{vpgfd} shows up. For the radiative transition $P
\rightarrow \gamma\gamma^{*}$, the Feynman diagrams are displayed in
Fig.~\ref{pggfd}.

Since the expressions for the various decay widths, such as $V \to P \gamma$,
$P \to V \gamma$ and $P \to \gamma \gamma$, are rather lengthy, we relegate them in Appendix~\ref{appendix-vpg-amp}.
The form factors of  ${F}_{P \rightarrow \gamma\gamma^{*}}(s)$ and ${F}_{V \rightarrow P \gamma^{*}}(s)$
can be found in Appendix~\ref{appendix-vpgstar-ff}.
The kinematics and amplitudes of $ V \to P \gamma^{*} \to P l^- l^+ $, $ P \to V \gamma^{*} \to V l^- l^+ $ and $P \to
\gamma\gamma^{*} \to \gamma l^- l^+ $, with the lepton $l=e, \mu$ are given in Appendix~\ref{appendix-vpll}.

\section{Short distance constraints from QCD}\label{he-constraint}

Given R$\chi$T being the dual theory of QCD in the resonance region,
the couplings appearing in the resonance chiral Lagrangian can not
be completely free, since there is only one parameter in QCD:
$\Lambda_{\rm QCD}$ and the heavy quark masses. To fix the resonance couplings in terms of
them requires solving the nonperturbative dynamics of
QCD from the first principle, which is exactly the reason that the
chiral effective field theory arises. Nevertheless, in the last
decades, a sufficient way to implement the short distance feature of
QCD has been developed, that is to match the operator product
expansion(OPE) of Green functions of QCD currents which are of order
parameter of the chiral symmetry breaking to the same quantity
calculated within the resonance chiral
Lagrangian~\cite{gf-moussallam-95,gf-moussallam-97,gf-bijnens,gf-knecht,
gf-pich-03}. Through this procedure, one could constrain
sufficiently the resonance couplings in certain cases.

The three-point Green function with vector-vector-pseudoscalar QCD
currents has been studied in different works.
Ref.~\cite{gf-moussallam-95} is devoted to the analysis of $P \to
\gamma \gamma$ by taking into the higher order Goldstone chiral
Lagrangian. The focus of Ref.~\cite{gf-knecht} is to estimate the
resonance contributions to the $\cO({p^6})$ LECs of $\chi$PT by
using the vector formalism to describe the vector resonances,
while Ref.~\cite{kampf11prd} exploits the antisymmetric tensor formalism
to incorporate the vector resonances.
In Ref.~\cite{gf-pich-03}, the $VVP$ Green function has been analyzed
by using the antisymmetric tensor representation for the vector
resonances and the phenomenology study focused on the radiative
processes of  $\omega \to \pi \gamma$ and $\pi \to \gamma\gamma$.
Our discussion in this section is devoted to the generalized study
of $VVP$ Green function given in~\cite{gf-pich-03} by extending the
content of pseudo-Goldstone bosons from the octet case to the nonet one, 
i.e. $u(x) \in SU(3) \to \tilde{u}(x) \in U(3) $. 

The $VVP$ Green function is defined as
\begin{eqnarray}
\int d^4x\int d^4ye^{i(p\cdot
x+q\cdot y )} \langle 0|T[V_{\mu}^{a}(x)V_{\nu}^{b}(y)P^c(0)]|0 \rangle
=d^{abc}\epsilon_{\mu\nu\alpha\beta}p^{\alpha}q^{\beta}{\Pi}_{VVP}(p^2,q^2,r^2)\,,
\end{eqnarray}
where the flavor indices are $a,b,c=0,...,8$; $r_\mu$ is defined as $r_{\mu}=-(p+q)_{\mu}$;
the vector and pseudoscalar currents are given by
\begin{eqnarray}
V_{\mu}^{a}(x)=(\bar{\psi}\gamma_{\mu}\frac{\lambda^a}{2}\psi)(x)\,, \quad
P^{a}(x)=(\bar{\psi}i\gamma_{5}\frac{\lambda^a}{2}\psi)(x)\,.
\end{eqnarray}
Taking $c=0$, one can get the singlet pseudoscalar current and $\Pi_{VVP}^{(0)}$
can be defined correspondingly. Although the evaluation of the singlet $\Pi_{VVP}^{(0)}$
and the octet  $\Pi_{VVP}^{(8)}$ in the intermediate energy region could lead to different results due to the $U_A(1)$ anomaly,
their asymptotic behaviors coincide in the chiral limit and at leading order of $1/N_C$~\cite{gf-moussallam-95}
\bqa\label{opevvp}
&&\lim_{\lambda\rightarrow \infty} {\Pi}_{VVP}^{(8)}[(\lambda p)^2,(\lambda
q)^2,(\lambda p+\lambda
q)^2]=\lim_{\lambda\rightarrow \infty} {\Pi}_{VVP}^{(0)}[(\lambda p)^2,(\lambda
q)^2,(\lambda p+\lambda
q)^2]= \nonumber \\ &&
-\frac{\langle \overline{\psi}\psi\rangle_0}{2\lambda^4}\frac{p^2+q^2+r^2}{p^2q^2r^2}[ 1 + \cO({\alpha_S}) ]+O(\frac{1}{\lambda^6})\,.
\eqa

Next let us focus on the calculation of the $VVP$ Green function within R$\chi$T. Since the new operators
given in Eq.(\ref{defc8d5}) only contribute to the singlet $\Pi_{VVP}^{(0)}$, the matching of the octet Green function
$\Pi_{VVP}^{(8)}$ does not lead to any new constraints on the resonance couplings, comparing with the results in~\cite{gf-pich-03}.
So we concentrate on the evaluation of the singlet $VVP$ Green function $\Pi_{VVP}^{(0)}$ in R$\chi$T and the result is found to be
\begin{eqnarray}\label{rxtvvp0}
 &&{\Pi}_{VVP}^{(0)}( p^2, q^2,r^2)=
-\frac{\langle \bar{\psi}\psi \rangle_0}{F^2} \bigg\{
-4F_V^2\frac{(\tilde{d}_1-\tilde{d}_3 )r^2 +
\tilde{d}_3(p^2+q^2)}{(M_V^2-p^2)(M_V^2-q^2)(M_0^2- r^2)} \nonumber
\\ && +2\sqrt2 \frac{F_V}{M_V}\frac{
(\tilde{c}_1+\tilde{c}_2-\tilde{c}_5) r^2+(\tilde{c}_2 +\tilde{c}_5
-\tilde{c}_1-2\tilde{c}_6)p^2
+(\tilde{c}_1-\tilde{c}_2+\tilde{c}_5)q^2   }{(M_0^2-
r^2)(M_V^2-p^2)} \nonumber \\&& +2\sqrt2 \frac{F_V}{M_V}\frac{
(\tilde{c}_1+\tilde{c}_2-\tilde{c}_5) r^2+(\tilde{c}_2 +\tilde{c}_5
-\tilde{c}_1-2\tilde{c}_6)q^2
+(\tilde{c}_1-\tilde{c}_2+\tilde{c}_5)p^2  }{(M_0^2-
r^2)(M_V^2-q^2)} \nonumber \\&& +\frac{32F_V^2\tilde{d}_2
}{(M_V^2-p^2)(M_V^2-q^2)} - \frac{16\sqrt2 F_V\tilde{c}_3 }{M_V}
\big( \frac{1}{M_V^2-p^2} +\frac{1}{M_V^2-q^2} \big)
+\frac{N_C}{8\pi^2 (M_0^2- r^2)} \nonumber \\&& +
4\sqrt{3}\,\tilde{c}_8M_VF_V[\frac{1}{(M_V^2-q^2)(M_0^2- r^2)}
+\frac{1}{(M_V^2-p^2)(M_0^2- r^2)}] \nonumber \\&& -
2\sqrt{6}\frac{{\tilde{d}_5}F_V^2M_V^2}{(M_V^2-p^2)(M_V^2-q^2)(M_0^2-
r^2)} \bigg\}\,.
\end{eqnarray}
We stress that the above expression for $\Pi_{VVP}^{(0)}$ is worked
out in the chiral limit. Due to the $U_A(1)$ anomaly, the singlet
pseudoscalar $\eta_1$ gains the non-vanishing mass $M_0$ even at the
chiral limit. Although the $\eta_1$ mass $M_0$ is suppressed by
$1/N_C$, its value is not a small quantity, which is even higher than
the lowest vector resonance mass $M_V$~\cite{Feldmann00}.
Hence to take this effect into account we have introduced the
non-vanishing mass for $\eta_1$ in the calculation of the
$\Pi_{VVP}^{(0)}$ function within R$\chi$T.

Matching the result evaluated from OPE of the singlet $VVP$ Green function displayed in Eq.(\ref{opevvp}) to the same quantity
evaluated within R$\chi$T, which is given in Eq.(\ref{rxtvvp0}), leads to the following constraints
\begin{eqnarray} \label{he-ope}
4\tilde{c}_3+\tilde{c}_1&=&0, \label{he-ope-c1}\\
\tilde{c}_1-\tilde{c}_2+\tilde{c}_5&=&0, \label{he-ope-c2}\\
\tilde{c}_5-\tilde{c}_6&=&\frac{N_C}{64\pi^2}\frac{M_V}{\sqrt{2}F_V}, \label{he-ope-c5}\\
\tilde{d}_1+8\tilde{d}_2 -\tilde{d}_3 &=& \frac{F^2}{8F_V^2}, \label{he-ope-d1}\\
\tilde{d}_3&=&-\frac{N_C}{64\pi^2}\frac{M_V^2}{F_V^2}+\frac{F^2}{8F_V^2} - \frac{\sqrt3 M_V}{F_V} \tilde{c}_8
- \frac{\sqrt2 M_0^2}{M_V F_V} \tilde{c}_1 \,.
\end{eqnarray}
We find the above constraints are all consistent with the ones given in~\cite{gf-pich-03}, except for the relation of $\tilde{d}_3$.
The consistent condition for the results of $\tilde{d}_3$ from the octet and singlet cases requires
\be\label{c8-constrain}
\tilde{c}_8 =  -\frac{ \sqrt2 M_0^2}{\sqrt3 M_V^2} \tilde{c}_1\,.
\ee
So in the numerical discussion, we will take this constraint for $ \tilde{c}_8 $.

There is another well known approach to address the
features of form factors corresponding to  exclusive
processes of QCD with high momenta transfer, which was developed
within the parton description scheme for the hadrons in Ref.~\cite{Lepage}.
The relevant one to our present discussion is the photon meson transition
form factor $F_{\pi\gamma}(Q^2)$, which can match the form factor defined in Eq.(\ref{defFF})
by applying the time reversal and replacing the vector resonance with the on-shell photon.
Although different approaches have predicted different asymptotic 
behaviors of $F_{\pi\gamma}(Q^2)$ at the order of $Q^{-2}$, for example
\bqa \label{he-ff-pg}
F_{\pi\gamma}(Q^2 \to \infty) &=& -\frac{F}{Q^2} \,,  \nonumber \\
F_{\pi\gamma}(Q^2 \to \infty)  &= & -\frac{2F}{3Q^2}\,,  \nonumber \\
F_{\pi\gamma}(Q^2 \to \infty)  &=&   -\frac{F}{3Q^2}\,,
\eqa
which are noted in Refs~\cite{Lepage}\cite{Manohar:1990}\cite{Gerard:1995} respectively,
they agree at the order of $ Q^0$, indicating the form factor behaving smoothly
at the high momentum transfer, i.e. $F_{\pi\gamma}(Q^2 \to \infty) \to 0$. So the most conservative
constraint would be just to impose the vanishing condition for the constant term in the form
factor, i.e. to demand the coefficient of $Q^0$ being zero. The explicit expression for the form factor
of $\pi\gamma\gamma^*$ can be found in Appendix~\ref{appendix-vpll} and the corresponding high energy
constraint from the order of $Q^0$ is
\begin{eqnarray}\label{he-ff-8-a}
\tilde{c}_1-\tilde{c}_2+\tilde{c}_5&=&0,\nonumber\\
\tilde{c}_5-\tilde{c}_6&=& \frac{F_V}{\sqrt{2}M_V} \tilde{d}_3 + \frac{N_C M_V}{32\sqrt2 \pi^2 F_V}\,,
\end{eqnarray}
and the corresponding constraint from the form factor of  $\omega\pi\gamma^*$, which is also given
in Appendix~\ref{appendix-vpll}, leads to
\begin{eqnarray}\label{he-ff-8-b}
\tilde{c}_1-\tilde{c}_2+\tilde{c}_5&=&0 \,,\nonumber\\
\tilde{c}_5-\tilde{c}_6&=& -\frac{F_V}{\sqrt{2}M_V} \tilde{d}_3\,.
\end{eqnarray}
Combining Eq.(\ref{he-ff-8-a}) and Eq.(\ref{he-ff-8-b}), we have the following relations
\begin{eqnarray}\label{he-ff-8}
\tilde{c}_1-\tilde{c}_2+\tilde{c}_5&=&0 \,,\nonumber\\
\tilde{c}_5-\tilde{c}_6&=&\frac{N_C}{64\pi^2}\frac{M_V}{\sqrt{2}F_V}\,,\nonumber\\
\tilde{d}_3&=&-\frac{N_C}{64\pi^2}\frac{M_V^2}{F_V^2}\,.
\end{eqnarray}

When discussing the high energy constraints, we take the chiral limit and $U(3)$ symmetry for the vector
resonances. In this case, the physical states $\eta$ and $\eta'$ are represented by
the flavour eigenstates $\eta_8$ and $\eta_1$, respectively.
So in the chiral limit, the constraints from the other processes with octet pseudoscalar mesons
lead to the same results, while the process involving the singlet pseudoscalar $\eta_1$ could lead to
different results as we keep the $U_A(1)$ anomaly effect, i.e. the non-vanishing
$\eta_1$ mass, in the deriving the high energy constraint.
The explicit result from the analysis of the $\eta_1 \gamma\gamma^*$ form factor is
\begin{eqnarray}\label{he-ff-1}
\tilde{c}_1-\tilde{c}_2+\tilde{c}_5&=&0,\nonumber\\
\tilde{c}_5-\tilde{c}_6&=&\frac{N_C}{64\pi^2}\frac{M_V}{\sqrt{2}F_V},\nonumber\\
\tilde{d}_3&=&-\frac{N_C}{64\pi^2}\frac{M_V^2}{F_V^2}- \frac{\sqrt3 M_V}{F_V} \tilde{c}_8
- \frac{\sqrt2 M_0^2}{M_V F_V} \tilde{c}_1 \,,
\end{eqnarray}
To demand the consistency of the results from $\pi\gamma\gamma^*$ and $\eta_1\gamma\gamma^*$,
we arrive at the same constraint in Eq.(\ref{c8-constrain}).

Notice that the only inconsistency from the OPE and form factors is
the relation of $\tilde{d}_3$, although fewer constraints are
obtained in the analysis of form factors. This observation has been
confirmed in other processes~\cite{gomez-dumm-10,guo-10}. The fact
that only one multiplet of resonances was unable to fulfill all the
high energy constraints from OPE and the corresponding form factors
was already noticed~\cite{gf-knecht,gf-bijnens,Mateu}. In this work,
we will take the result for $\tilde{d}_3$ in Eq.(\ref{he-ff-8})
obtained from the high energy constraint of form factor, which is
more related to the processes we are discussing. To reduce the free
parameters as many as possible in the phenomenology discussion, we
also exploit the constraints in Eq.(\ref{he-ope-c1}) and
Eq.(\ref{he-ope-d1}) obtained from the OPE analysis. In summary, the
high energy constraints we are going to use in the following
discussion are those in Eq.(\ref{he-ope-c1}), Eq.(\ref{he-ope-d1}),
Eq.(\ref{c8-constrain}) and Eq.(\ref{he-ff-8}).

\section{ Phenomenology discussion } \label{pheno}

Although we can fix some parameters through the short distance
constraints from QCD in the previous section, some of the resonance
couplings appearing in the decay widths given in Appendix are still
unconstrained. So we need to fit the unknown resonance couplings,
such as $\tilde{c}_3, \tilde{d}_2$ and $\tilde{d}_5$, together with
the $\eta-\eta'$ mixing parameters $\theta_0, \theta_8, F_0$ and
$F_8$. For the mass of vector resonances in the chiral limit $M_V$,
one can safely estimate its value by $M_\rho$, the mass of
$\rho(770)$~\cite{guo-09}. While for the parameter $F_V$, describing
the transition strength of the neutral vector resonances and photon,
its value is still a somewhat controversial subject and several
solutions have been proposed in different works. In the pioneer work
to discuss the high energy constraints in R$\chi$T~\cite{Ecker89},
$F_V= \sqrt2 F$ was predicted by combining the high energy
constraints from the pion vector and axial-vector form factors
within the minimal R$\chi$T Lagrangian at leading order of $1/N_C$,
while in the next-to-leading order analysis of the pion vector form
factor $F_V= \sqrt3 F$ is updated in Ref.~\cite{pich-10-vff}, which
has also been confirmed in the study of radiative tau
decays~\cite{guo-10} and the partial wave analysis of $\pi\pi$
scattering~\cite{guo-07-jhep,guo11prd}. Phenomenology determinations of $F_V
= 147$ MeV and 180 MeV have been used in $\tau \to V P
\nu_\tau$~\cite{guo-08} and  $\tau \to K \bar{K}
\pi$~\cite{gomez-dumm-10} decays respectively. By estimating the
pseudo-Goldstone decay constant at chiral limit by the pion decay
constant $F_\pi$, one has
\bqa
F_V = \sqrt2 F = 131\,{\rm MeV}\,,
\qquad F_V = \sqrt3 F = 160\,{\rm MeV}\,.
\eqa
Apparently, a more precise value for $F_V$ is needed in our discussion, since the
physical processes we are discussing are mainly the radiative decays
of vector resonances and precisely $F_V$ describes the interaction
of vector resonances and photon. Thus we decide to free $F_V$ and
fit its value in our program, in such a way we could predict a more
reliable value, as $F_V$ is rather sensitive to the processes we are
considering. To exploit the high energy constraint for $\tilde{c}_8$
in Eq.(\ref{c8-constrain}), the value for the $U_A(1)$ anomaly mass
$M_0$ is needed, which has been reviewed in Ref.~\cite{Feldmann00} and also
recently determined in Ref.~\cite{guo11prd}.
We use the average value $M_0 = 900$ MeV from the two mentioned references throughout.
For the other unmentioned inputs,
unless an explicit statement is given, we will take the
corresponding values from Ref.~\cite{pdg}.

\begin{figure}[H]
\includegraphics[angle=0,width=1.0\textwidth]{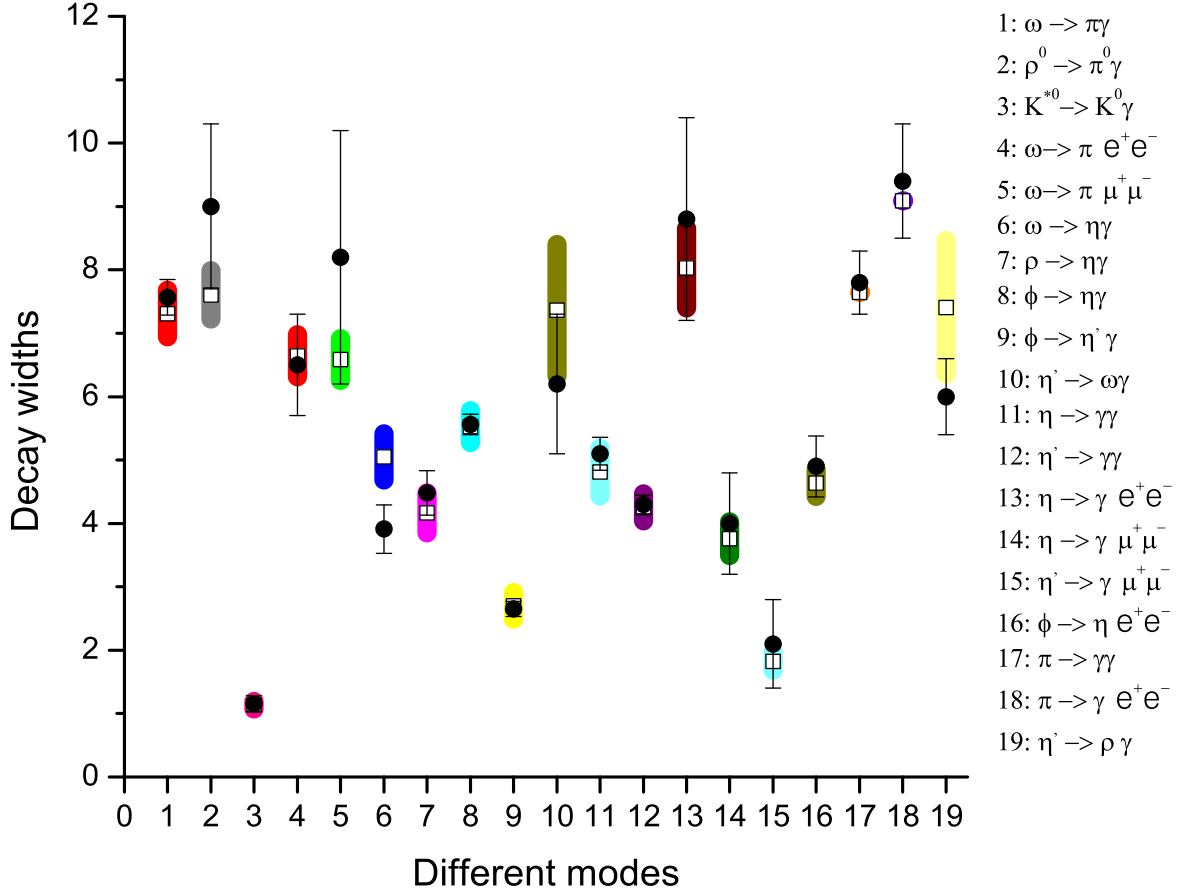}
\caption{ Different decay widths. We have scaled different decay
widths into a common region in order to show them in one figure. For
the values before scaling, see the numbers in
Tables~\ref{table-noeta} and \ref{table-eta}. The open squares
denote the central values with the best fit given in
Eq.\eqref{fit-result}, and the shaded area correspond to the error
bands generated by the parameter configurations explained in the
text after Eq.\eqref{fit-result}. Note that using the high energy
constraints we can completely predict the decay widths of
$\pi\rightarrow\gamma\gamma$ and $\pi\rightarrow\gamma e^- e^+$,
which are in good agreement with the experimental data, and here we
include these two processes just for completeness. We do not include
$\eta' \to \rho \gamma$ in the fit, since PDG~\cite{pdg} also
includes the background part from $\eta' \to \pi\pi \gamma$ to
determine the width for $\eta' \to \rho \gamma$. Nevertheless due to
the dominant decay channel of the $\rho$ resonance is $\pi\pi$, our
prediction for $\eta' \to \rho \gamma$ agrees with the one
from PDG~\cite{pdg}. }\label{fig.decaywidths}
\end{figure}

\begin{figure}[ht]
\includegraphics[angle=0,width=0.9\textwidth]{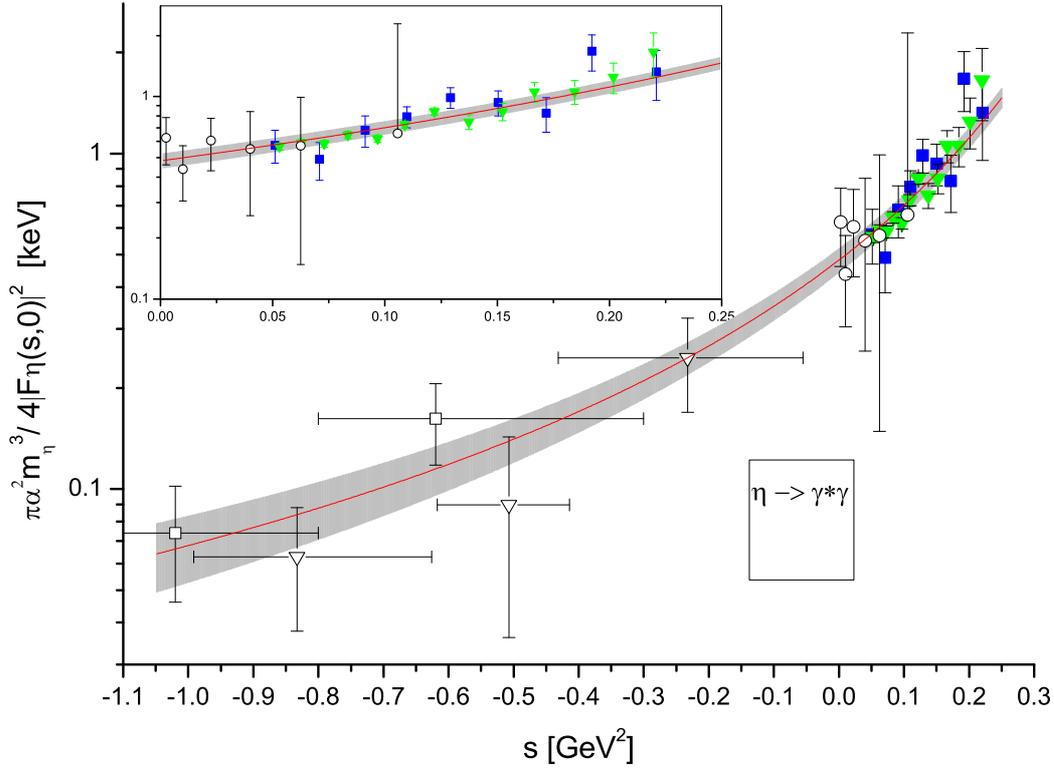}
\caption{ The form factors of $ \eta\rightarrow \gamma^{*}\gamma$.
The solid line (red) denotes the result from the best fit and the shaded area correspond to the error bands.
Sources of the different experiment data are:
solid squares ~\cite{LG1,LG2}, open squares ~\cite{CELLO}, open
circles ~\cite{SND}, solid triangles~\cite{NA60}, open
triangles~\cite{TPC}. The separated figure in the upper part is the close-up of the main plot in the
region of $s>0$.  \label{fig.etaff} }
\end{figure}

\begin{figure}[ht]
\includegraphics[angle=0,width=0.9\textwidth]{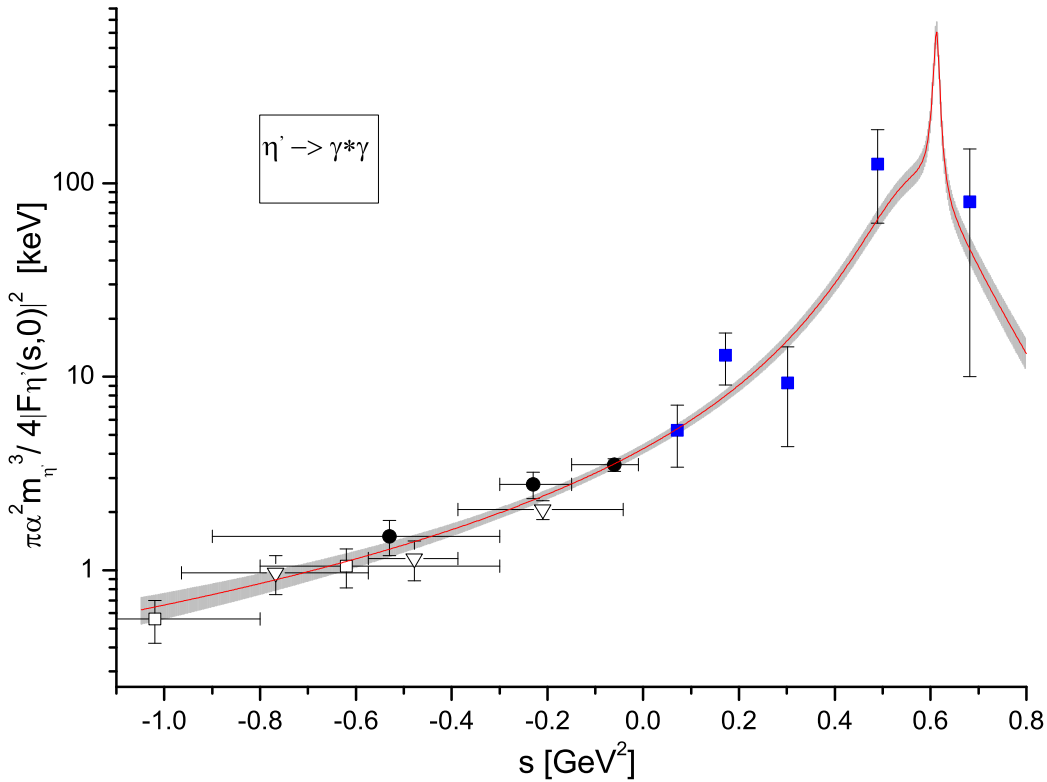}
\caption{ The form factors of $ \eta'\rightarrow
\gamma^{*}\gamma$. The solid line (red) denotes the result from the best fit
and the shaded area correspond to the error bands. Sources of the different experiment data are:
solid squares ~\cite{LG1,LG2}, open squares
~\cite{CELLO}, open triangles~\cite{TPC}, solid circles ~\cite{L3}.
\label{fig.etapff}}
\end{figure}

\begin{figure}[ht]
\includegraphics[angle=0,width=0.9\textwidth]{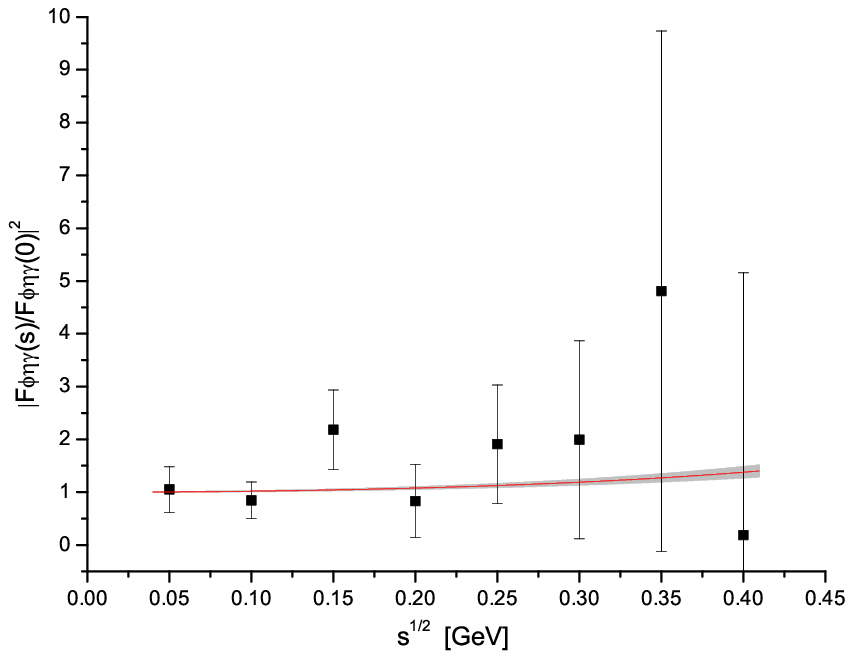}
\caption{ The form factors of $\phi\rightarrow
\eta\gamma^{*}$~\cite{SND}. The solid line (red) denotes the result from the best fit
and the shaded area correspond to the error bands.  \label{fig.phietaff} }
\end{figure}

As advertised previously, although the main contribution of our
current work is to determine the $\eta-\eta'$ mixing parameters in a
more reliable theoretical framework, we include the relevant
radiative processes without $\eta$ and $\eta'$ into our discussion
as well. By performing the $\chi^2$ fit, we can determine the unknown
resonance couplings and  $\eta-\eta'$ mixing parameters:
\begin{eqnarray} \label{fit-result}
&&F_8 =(1.37\pm 0.07) F_{\pi}\,,\qquad F_0 =(1.19\pm 0.18)
F_{\pi}\,,\nonumber\\ && \theta_8=(-21.1\pm 6.0 )^{\circ}\,, \qquad
\,\,\, \theta_0=(-2.5\pm 8.2 )^{\circ}\,,\nonumber\\ &&
F_V=(136.6\pm 3.5) {\rm MeV} \,, \qquad \tilde{c}_3=0.011\pm 0.016
\,, \nonumber\\ && \tilde{d}_2=0.086\pm0.085\,,  \qquad \qquad
\tilde{d}_5= 0.36 \pm 0.40 \,,
\end{eqnarray}
with $\chi^2/{\rm d.o.f }= 64.0/(70-8)=1.03$.
For the various decay widths, we summarize the experiment data and
the results from our fitting program in Table~\ref{table-noeta} for
the processes without $\eta$ and $\eta'$ and in
Table~\ref{table-eta} for those involving $\eta$ or $\eta'$.
To visualize the results, we plot the numbers in Tables~\ref{table-noeta}
and ~\ref{table-eta} in Fig.~\ref{fig.decaywidths}, where one should notice
we have scaled different decay widths to a proper range in order to
show them in one figure. The resulting plots for the form factors of $\eta\gamma\gamma^*$,
$\eta'\gamma\gamma^*$ and $\phi\eta\gamma^*$ are given in Figs.~
\ref{fig.etaff}, \ref{fig.etapff} and \ref{fig.phietaff}, respectively.
The error bands shown in the plots and the errors of the parameters in Eq.~(\ref{fit-result})
correspond to the statistical uncertainties at 2 standard deviations~\cite{etkin82prd}:
$n_\sigma = (\chi^2 - \chi_0^2)/\sqrt{2\chi_0^2}$, with $\chi_0^2$ the minimum $\chi^2$
obtained in the fit and $n_\sigma$ the number of standard deviations.

\begin{table}[ht]
\begin{small}
\begin{center}
\begin{tabular}{cccccc}\hline\hline
&Exp&         Fit & Theo {\scriptsize($F_V=160$ MeV)}\,&\, Theo {\scriptsize($F_V=180$ MeV)}    \\
\hline
$\Gamma_{\omega\rightarrow\pi\gamma}$ &  $757\pm28$ & $731\pm37$&533&421 \\
$\Gamma_{\rho^0\rightarrow\pi^0\gamma}$ &  $89.6\pm12.6$  & $76.0\pm38$ &55.4&43.8 \\
$\Gamma_{K^{*0}\rightarrow K^0\gamma}$ &   $116\pm12$  & $113\pm6$ &83&65 \\
$\Gamma_{\omega \rightarrow\pi e^- e^+}$ &   $6.54\pm0.83$  & $6.64\pm0.33$&4.84&3.83  \\
$\Gamma_{\omega \rightarrow\pi \mu^- \mu^+}$ & $0.82\pm0.21$  & $0.66\pm0.03$ &0.48&0.38 \\
\hline\hline
\end{tabular}
\caption{\label{table-noeta} Experimental and theoretical values of
the various decay widths without $\eta$ and $\eta'$. The experiment data are taken from ~\cite{pdg}. All of the values 
are given in units of KeV unless  specified. The results from
our fit are listed in the column Fit and the error bands of the widths
are calculated by using the same parameter configurations that
we use to get the error bands for the parameters in Eq.~\eqref{fit-result}.
To show the relevance of $F_V$, we have given another two theoretical predictions for the
various decay widths in the last two columns by taking $F_V = 160,
180$ MeV.
 }
\end{center}
\end{small}
\end{table}

\begin{table}[ht]
\begin{small}
\begin{center}
\begin{tabular}{cccc}\hline\hline
&Exp &           Fit    \\
\hline
$\Gamma_{\omega\rightarrow\eta\gamma}$ & $3.91\pm 0.38$  &  $5.05\pm 0.36$  \\
$\Gamma_{\rho^0\rightarrow\eta\gamma}$ & $44.8\pm 3.5$     & $41.6\pm 3.2$ \\
$\Gamma_{\phi\rightarrow \eta\gamma}$ & $55.6\pm 1.6$   &  $55.3\pm 2.5$  \\
$\Gamma_{\phi\rightarrow \eta'\gamma}$ & $0.265\pm 0.012$   &  $0.270\pm0.021$  \\
$\Gamma_{\eta'\rightarrow\omega\gamma}$ & $6.2\pm 1.1 $ \,\,  & \,\,  $7.4\pm 1.0$  \\
$\Gamma_{\eta\rightarrow\gamma\gamma}$ & $0.510\pm 0.026 $   \,\,  & \,\, $0.481\pm 0.038$  \\
$\Gamma_{\eta'\rightarrow\gamma\gamma}$ & $4.30\pm 0.15$   \,\,  & \,\, $4.25\pm 0.21$  \\
$\Gamma_{\eta\rightarrow\gamma e^- e^+}$ & $(8.8\pm 1.6)\times 10^{-3}$   \,\,  & \,\, $(8.0\pm 0.6)\times 10^{-3}$  \\
$\Gamma_{\eta\rightarrow\gamma \mu^- \mu^+}$ & $(0.40\pm 0.08)\times 10^{-3}$   \,\,  & \,\, $(0.38\pm 0.03)\times 10^{-3}$  \\
$\Gamma_{\eta'\rightarrow\gamma \mu^- \mu^+}$ & $(2.1\pm 0.7)\times 10^{-2}$   \,\,  & \,\,$(1.8\pm 0.1)\times 10^{-2}$  \\
$\Gamma_{\phi \rightarrow\eta e^- e^+}$ & $ 0.490\pm 0.048$  \,\,  & \,\,  $0.464\pm 0.021$  \\
\hline\hline
\end{tabular}
\caption{\label{table-eta}  Experimental and theoretical values of
the various decay widths involving $\eta$ and $\eta'$. 
The experiment data are taken from ~\cite{pdg}. All of the values are given 
in units of KeV unless specified.  The error bands
of the widths are calculated by using the same parameter configurations 
as used in Table \ref{table-noeta}. }
\end{center}
\end{small}
\end{table}

Several remarks about the fitting results are in order. We comment
them as follows.
\begin{enumerate}
\item
The first lesson we can learn from the results in Eq.(\ref{fit-result}) is that
$\tilde{c}_3$, $\tilde{d}_2$ and $\tilde{d}_5$ carry huge error bars.
Nevertheless we find there exist strong correlations among these parameters.
We plot the correlations of $\tilde{d}_2$-$\tilde{d}_5$, $\tilde{c}_3$-$\tilde{d}_2$
and $\tilde{c}_3$-$\tilde{d}_5$ respectively in Figs.(\ref{fig.d2d5correlations}),
(\ref{fig.c3d2correlations}) and (\ref{fig.c3d5correlations}),
where the same parameter configurations have been used as we exploit
in evaluating the error bands in Eq.(\ref{fit-result}) and
Figs.(\ref{fig.etaff}-\ref{fig.phietaff}).
A very strong linear correlation between $\tilde{d}_2$ and $\tilde{d}_5$ is observed,
as one can see in Fig.(\ref{fig.d2d5correlations}).
While the correlations of $\tilde{c}_3$-$\tilde{d}_2$ and $\tilde{c}_3$-$\tilde{d}_5$,
as shown in Figs.(\ref{fig.c3d2correlations}) and (\ref{fig.c3d5correlations}),
are not as strong as $\tilde{d}_2$-$\tilde{d}_5$.
A correlation between $\tilde{c}_3$ and $\tilde{d}_2$ (in fact they are $c_3$ and $d_2$
from the pseudo-Goldstone octet Lagrangian)
has been revealed in a preliminary analysis of $\tau \to \pi\pi\eta \nu_\tau$ decay~\cite{roig10npbps},
where $\eta$ particle is treated as the pure octet $\eta_8$.
The parameter space found in the previous reference covers most of the space
shown in Fig.~(\ref{fig.c3d2correlations}), but the correlation relation in ~\cite{roig10npbps}
is with opposite sign of the relation we find in the present work.
This indicates that the combining study of the radiative decay processes
of $\eta$ or $\eta'$ and the $\tau$ decays involving $\eta$ or $\eta'$
may help us pin down the resonance parameters,
such as $\tilde{c}_3$ and $\tilde{d}_2$, which deserves a future work.
For the remaining parameters in Eq.(\ref{fit-result}), we do not find significant
correlations among them.

\begin{figure}[ht]
\includegraphics[angle=0,width=0.9\textwidth]{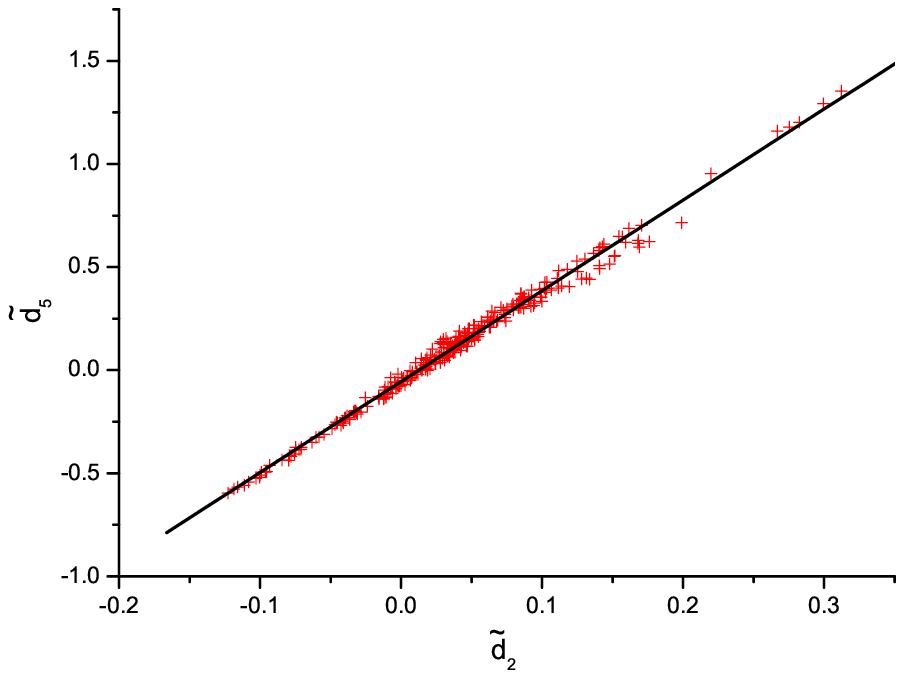}
\caption{ The correlations between $\tilde{d}_2$ and $\tilde{d}_5$. 
The solid line (black) corresponds to $\tilde{d}_5=4.4\tilde{d}_2 - 0.06$.
}\label{fig.d2d5correlations}
\end{figure}

\begin{figure}[ht]
\includegraphics[angle=0,width=0.9\textwidth]{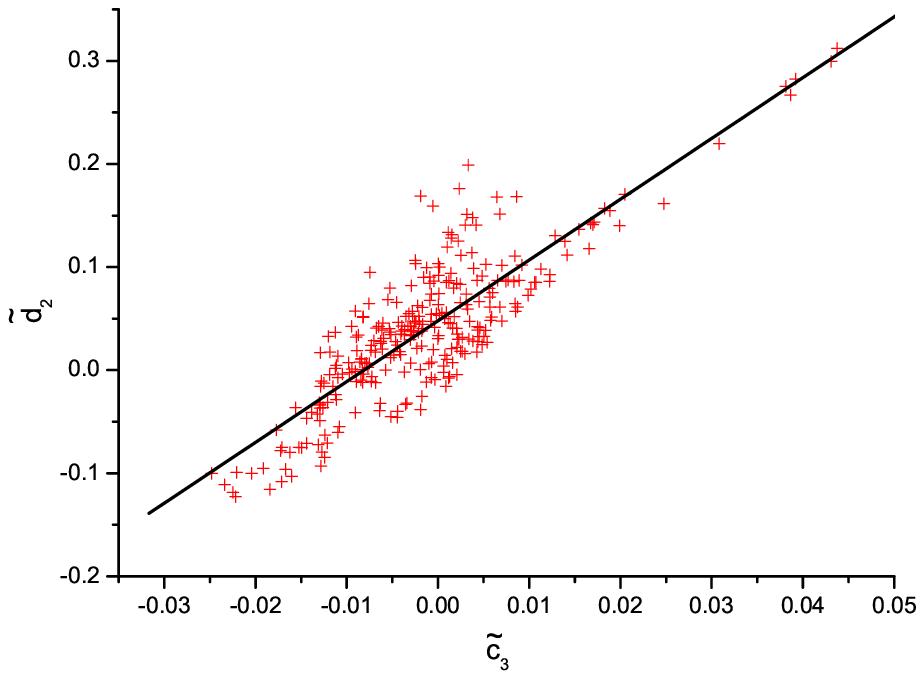}
\caption{ The correlations between $\tilde{c}_3$ and $\tilde{d}_2$.
The solid line (black) corresponds to $\tilde{d}_2= 5.6\tilde{c}_3 + 0.06$.
}\label{fig.c3d2correlations}
\end{figure}

\begin{figure}[ht]
\includegraphics[angle=0,width=0.9\textwidth]{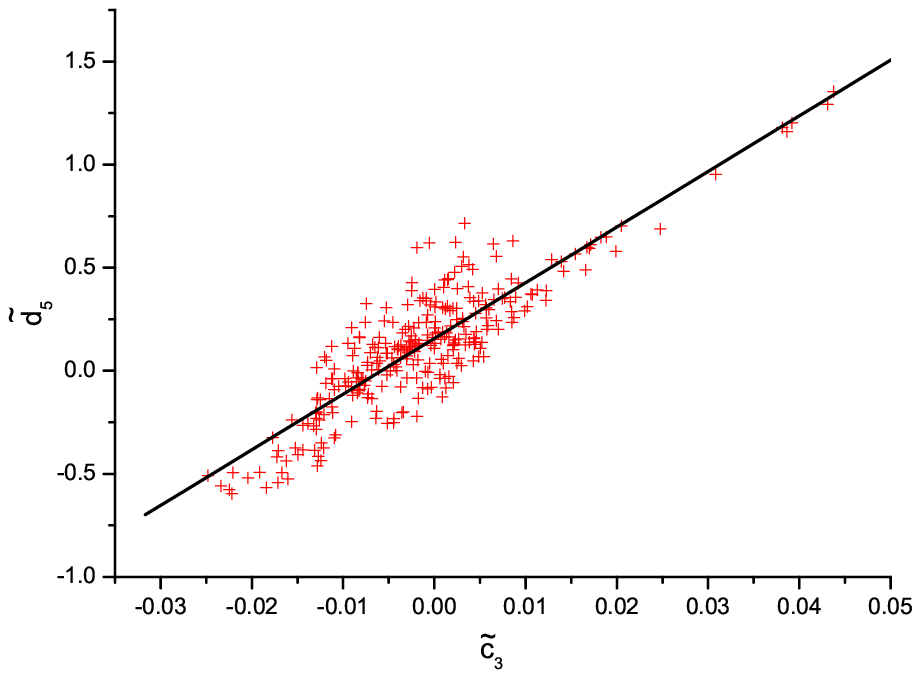}
\caption{ The correlations between $\tilde{c}_3$ and $\tilde{d}_5$.
The solid line (black) corresponds to $\tilde{d}_5=25\tilde{c}_3 +0.2$. 
}\label{fig.c3d5correlations}
\end{figure}

\item The value of $F_V$. As one can see in Eq.(\ref{fit-result}),
our analysis favors a smaller value for $F_V$,
comparing with the values used in \cite{guo-08} and \cite{gomez-dumm-10}.
By using the high energy constraints mentioned in previous section,
the processes appearing in Table~\ref{table-noeta} are solely determined by $F_V$. Thus
those radiative decay processes without $\eta$ and $\eta'$ provide a
tight constraint on the value of $F_V$. To clearly show the
relevance of $F_V$ in those channels, we give another two
predictions by taking $F_V = 160$ and $180$ MeV in the last two
columns in Table~\ref{table-noeta}. We also use these 5 processes in
Table~\ref{table-noeta} to perform a fit to get the value of $F_V$,
and the result is $F_V=(134.05\pm 2.17) {\rm MeV}$, with
$\chi^2/{\rm d.o.f }=\frac{1.46}{5-1}=0.36$.
This result for $F_V$ is in perfect agreement with the global fit in Eq.(\ref{fit-result}).
If one takes the value of $F_V=134.05 {\rm MeV}$ and fits the other 7 unknown parameters
by using the rest of experimental data as we used to get Eq.\eqref{fit-result},
the fit results turn out to be quite similar to the ones we show in Eq.\eqref{fit-result},
as expected.

\item The $\eta-\eta'$ mixing parameters. For $F_8$,
it can be completely fixed by the ratio of $F_K/F_\pi$ in the
next-to-next-to-leading order within the triple expansion of large
$N_C$ $\chi$PT, which yields a rather reliable prediction $F_8 =
1.34 F_\pi$~\cite{Kaiser98}. However at the same order, there exist
several unknown LECs for the predictions of $F_0$ and $\theta_0 -
\theta_8$, which prevents the precise predictions for their values.
As one can see in Eq.(\ref{fit-result}), our result for $F_8$ agrees
with the $\chi$PT prediction. In Ref.~\cite{Leutwyler98}, $F_0$ was
determined in the process $P\to\gamma\gamma$ at next-to-leading
order by ignoring the chiral symmetry breaking operators. Assuming
OZI violating coupling in the next-to-leading order to vanish, $F_0
= 1.25 F_\pi$ can be derived~\cite{Leutwyler98,Feldmann99}. $F_0$
can be also evaluated in the standard way to include the chiral
corrections from loops and LECs in the calculation of the
axial-vector current matrix element~\cite{Kaiser98}. In this case,
two additional OZI violating couplings appear and if one assumes
those couplings to vanish, $F_0 \simeq F_\pi$ can be predicted. So a
reliable determination of $F_0$ could help us better understand the
somewhat inconsistent results from the two approaches. As an
improvement, we have included not only the OZI violating operators
but also the chiral symmetry breaking ones in R$\chi$T to determine
the value of $F_0$ in $P\to \gamma\gamma$ processes. Our result
shows the inclusion of the OZI suppressed and higher chiral symmetry
breaking operators in the calculation of $P\to\gamma\gamma$ does not
change the result of $F_0 = 1.25 F_\pi$ very much, as obtained in
\cite{Leutwyler98}. This also indicates that the ignorance of the
OZI suppressed operators in the calculation of the axial-vector
current matrix element, which leads to $F_0 \simeq F_\pi$, is not a
good approximation. About the mixing angles, our result
$\theta_8=(-21.1\pm 6.0)^{\circ}$ agrees well with the 
results in literature, see Table 1 of Ref.~\cite{Feldmann00}. While
for the mixing angle $\theta_0$, our analysis reveals that a huge
error accompanies this parameter, as shown in Eq.\eqref{fit-result}.
This may be viewed as a source that why rather different results
have been obtained for $\theta_0$~\cite{Feldmann00}.

By using simple parameterizations of the VP$\gamma$ and
P$\gamma\gamma$ vertexes that only consist of constant terms
(independent of the quark masses and momenta), the $\eta-\eta'$
mixing parameters have been explored in various $V \to P \gamma$ and
$P \to \gamma \gamma$ processes and one can see
Ref.~\cite{Feldmann00} for a comprehensive analysis of different
results. Our current work confirms the validity of the
two-mixing-angle description in a more general framework for
VP$\gamma$ and P$\gamma\gamma$ interaction vertexes. Within the
current theoretical framework, the various data from different
radiative decay processes can be simultaneously incorporated and our
$\chi^2$ is clearly better than that in Ref.~\cite{Escribano05}.

\item An additional interest of our work is to test the resonance
saturation assumption for the couplings
relevant to the chiral anomaly induced processes $P\to \gamma
\gamma$, with $P= \pi, \eta, \eta'$. The next-to-leading order odd
intrinsic parity operators in the chiral Lagrangian with only pseudo
Goldstone bosons can be categorized into two parts: $O(N_C^1p^6)$
and $O(N_C^0p^4)$~\cite{gf-moussallam-95,bijnens-01}:
\begin{eqnarray} \label{lagp6wzw}
\tilde{L}_{odd}^{(2)}=i\tilde{t}_1\varepsilon_{\mu\nu\alpha\beta}
\langle
\tilde{\chi}_{-}\tilde{f}_{+}^{\mu\nu}\tilde{f}_{+}^{\alpha\beta}
\rangle -\tilde{t}_2\varepsilon_{\mu\nu\alpha\beta} \langle
\nabla_{\lambda}f_{+}^{\lambda\mu}\{f_{+}^{\alpha\beta}, u^{\nu}\}
\rangle +\tilde{k}_3\varepsilon_{\mu\nu\alpha\beta} \langle
\tilde{f}_{+}^{\mu\nu}\tilde{f}_{+}^{\alpha\beta} \rangle
\sqrt{6}\tilde{\phi}_0. \label{eqchiptt1k3}
\end{eqnarray}
The higher order LECs appearing in the pseudo-Goldstone chiral
Lagrangian encode the high energy dynamics of the underlying theory.
In the resonance saturation approach, it is assumed that the high
order LECs are completely saturated by the resonances and thus one
does not need to include extra pure higher order pseudo-Goldstone
operators in the resonance Lagrangian. This assumption has been
proven to be very successful for the $O(p^4)$ $\chi$PT LECs in the
even intrinsic parity sector~\cite{Ecker}. In the present
discussion, we shall check this assumption in the odd intrinsic
parity sector. By integrating out the vector resonances in the
resonance chiral Lagrangian introduced in Sect.~\ref{rxt-intro}, we
have the following predictions
\begin{eqnarray} \label{t1rxt}
\tilde{t}_1^V &=&-\frac{F_V}{4\sqrt{2}M_V^3}
(\tilde{c}_1+\tilde{c}_2+8\tilde{c}_3-\tilde{c}_5) +\frac{F_V^2}{8
M_V^4}(\tilde{d}_1+8\tilde{d}_2-\tilde{d}_3)\,,  
\\  \label{t2rxt}
\tilde{t}_2^V &=&-\frac{F_V}{\sqrt{2}M_V^3} (\tilde{c}_5-\tilde{c}_6)
+\frac{F_V^2}{2 M_V^4} \tilde{d}_3\,,
\\   \label{k3rxt}
\tilde{k}_3 &=& -\frac{F_VM_0^2}{6\sqrt{2}M_V^3}(\tilde{c}_1+\tilde{c}_2-\tilde{c}_5)
-\frac{F_V\tilde{c}_8}{2\sqrt{3}M_V} +\frac{F_V^2M_0^2}{12
M_V^4}(\tilde{d}_1-\tilde{d}_3)
+\frac{F_V^2\tilde{d}_5}{2\sqrt{6}M_V^2}\,,
\end{eqnarray}
where $\tilde{t}_1^V, \tilde{t}_2^V$ coincide with the results in
Ref.~\cite{gf-pich-03} and the result for $\tilde{k}_3$ is a new
result to our knowledge. Taking into account the high energy
constraints of the resonance couplings in
Eqs.(\ref{he-ope-c1})(\ref{he-ope-d1})(\ref{c8-constrain})
and (\ref{he-ff-8}), we have the simplified predictions
\begin{eqnarray}\label{t1t2k3rxt-he}
\tilde{t}_1^V = \frac{F^2}{64 M_V^4}\,,\quad \tilde{t}_2^V =
-\frac{N_C}{64\pi^2 M_V^2}\,, \quad  \tilde{k}_3^V = \frac{F^2
M_0^2}{96 M_V^4} -\frac{2 F_V^2 M_0^2 }{3 M_V^4} \tilde{d}_2
+\frac{F_V^2 }{2\sqrt6 M_V^2} \tilde{d}_5 \,,
\end{eqnarray}

The fitted results in Eq.(\ref{fit-result}) yield
\begin{eqnarray}
\tilde{t}_1= 0.37\times 10^{-3} \,{\rm GeV}^{-2}\,, \quad
\tilde{t}_2= -8.0\times 10^{-3} \,{\rm GeV}^{-2}\,, \quad
\tilde{k}_3= (0.83 \pm 3.02) \times 10^{-4}\,, \label{t1k3fit}
\end{eqnarray}
where the error of $k_3$ is estimated by taking the errors from the
fit results in Eq.(\ref{fit-result}).

Now we can calculate the form factors of $P \to \gamma \gamma$ using
the pseudo-Goldstone Lagrangian in Eq.(\ref{lagp4wzw}) and
Eq.(\ref{lagp6wzw}). For $\pi \to \gamma\gamma$, the $k_3$ operator
is irrelevant and our result is the same as the one in
Ref.~\cite{gf-pich-03} \bqa F_{\pi \rightarrow \gamma\gamma} =
F_{\pi \rightarrow \gamma\gamma}^{\rm WZW} + F_{\pi \rightarrow
\gamma\gamma}^{\tilde{t}_1} \,, \eqa where $F_{\pi \rightarrow
\gamma\gamma}^{\rm WZW} $ denotes the contribution from the WZW
Lagrangian in Eq.(\ref{lagp4wzw}) \be F_{\pi\rightarrow
\gamma\gamma}^{\rm WZW}=-\frac{1}{4\pi^2 F_\pi} \,, \ee and $F_{\pi
\rightarrow \gamma\gamma}^{\tilde{t}_1}$ denotes the contribution
from the $\tilde{t}_1$ term
\begin{eqnarray}
{F}_{\pi
\rightarrow\gamma\gamma}^{\tilde{t}_1}=\frac{64}{3F_\pi}m_{\pi}^2\tilde{t}_1\,.
\end{eqnarray}
Note that the $\tilde{t}_2$ operator does not contribute to the
considered process. For $\eta \to \gamma \gamma$, we have \bqa
F_{\eta \rightarrow \gamma\gamma} = F_{\eta \rightarrow
\gamma\gamma}^{\rm WZW} + F_{\eta \rightarrow
\gamma\gamma}^{\tilde{t}_1} + F_{\eta \rightarrow
\gamma\gamma}^{\tilde{k}_3}\,, \eqa where \be F_{\eta
\rightarrow\gamma\gamma}^{\rm WZW}=-\frac{1}{
\cos{(\theta_0-\theta_8)}}\frac{1}{4\sqrt{3}\pi^2}(\frac{\cos\theta_0}{F_8}
-\frac{\sqrt{8}\sin\theta_8}{F_0}) , \ee and $F_{\eta \rightarrow
\gamma\gamma}^{\tilde{t}_1},\, F_{\eta \rightarrow
\gamma\gamma}^{\tilde{k}_3}$ denote the contributions from
$\tilde{t}_1, \tilde{k}_3$ operators respectively
\begin{eqnarray}
{F}_{\eta \rightarrow \gamma\gamma}^{\tilde{t}_1}=\frac{1}{
\cos{(\theta_0-\theta_8)}}\{\frac{64}{9\sqrt{3}}(7m_\pi^2-4m_K^2)\frac{\cos\theta_0}{F_8}
+\frac{128\sqrt{2}}{9\sqrt{3}}(2m_\pi^2+m_K^2)\frac{-\sin\theta_8}{F_0}
 \}\tilde{t}_1,
\end{eqnarray}
\begin{eqnarray}
{F}_{\eta \rightarrow \gamma\gamma}^{\tilde{k}_3}=-\frac{1}{
\cos{(\theta_0-\theta_8)}}\frac{64\sqrt{6}\sin\theta_8}{3F_0}\tilde{k}_3\,.
\end{eqnarray}
Similar result for $\eta' \to \gamma\gamma$ is found to be \bqa
F_{\eta' \rightarrow \gamma\gamma} = F_{\eta' \rightarrow
\gamma\gamma}^{\rm WZW} + F_{\eta' \rightarrow
\gamma\gamma}^{\tilde{t}_1} + F_{\eta' \rightarrow
\gamma\gamma}^{\tilde{k}_3}\,, \eqa where \be F_{\eta' \rightarrow
\gamma\gamma}^{\rm WZW} =-\frac{1}{
\cos{(\theta_0-\theta_8)}}\frac{1}{4\sqrt{3}\pi^2}(\frac{\sin\theta_0}{F_8}
+\frac{\sqrt{8}\cos\theta_8}{F_0}), \ee
\begin{eqnarray}
{F}_{\eta' \rightarrow \gamma\gamma}^{\tilde{t}_1}=\frac{1}{
\cos{(\theta_0-\theta_8)}}\{\frac{64}{9\sqrt{3}}(7m_\pi^2-4m_K^2)\frac{\sin\theta_0}{F_8}
+\frac{128\sqrt{2}}{9\sqrt{3}}(2m_\pi^2+m_K^2)\frac{\cos\theta_8}{F_0}
 \}\tilde{t}_1,
\end{eqnarray}
\begin{eqnarray}
{F}_{\eta' \rightarrow \gamma\gamma}^{\tilde{k}_3}=\frac{1}{
\cos{(\theta_0-\theta_8)}}\frac{64\sqrt{6}\cos\theta_8}{3F_0}\tilde{k}_3.
\end{eqnarray}

In Table \ref{tableformfactorPgg}, we show different contributions
to the form factor ${F}_{P\rightarrow \gamma\gamma}$ using the
predictions for $\tilde{t}_1$ and $\tilde{k}_3$ in
Eq.(\ref{t1k3fit}). For the $\pi \to \gamma\gamma$ process, the
chiral symmetry breaking effect is rather tiny, about 1\%, since the
chiral correction is proportional to $m_\pi^2 / M_V^2$. Hence the
leading order contribution from the WZW Lagrangian overwhelmingly
dominates the decay width of $\pi \to
\gamma\gamma$~\cite{gf-pich-03,Goity}. For $\eta \to \gamma \gamma$
and $\eta' \to \gamma\gamma$, the WZW term can also give rather
close results to the experimental values, as one can see in Table
\ref{tableformfactorPgg} the next-to-leading order correction is
only at most 14\% of the WZW term. So our current calculations
confirm the validity of the triple expansion scheme for the odd
intrinsic parity pseudo-Goldstone Lagrangian, i.e. the WZW
contribution plays the dominant role in the $P \to \gamma\gamma$
processes. We point out that this conclusion is based on the fact
we have used the fitted results for the mixing parameters in
Eq.(\ref{fit-result}).

\begin{table}[ht]
\begin{center}
\begin{tabular}{|c||c|c|c|c|c|}
\hline
 &${F}_{P \rightarrow \gamma\gamma}^{WZW}$&${F}_{P \rightarrow \gamma\gamma}^{\tilde{t}_1}$ &
 ${F}_{P \rightarrow \gamma\gamma}^{\tilde{k}_3}$
     & ${F}_{P \rightarrow \gamma\gamma}^{WZW+\tilde{t}_1+\tilde{k}_3}$ & ${F}_{P \rightarrow \gamma\gamma}^{EX}$\\
\hline   $\pi \rightarrow \gamma\gamma$  & $-0.274$  &$ 0.002 $ & 0 & $-0.272 $& $ - 0.275\pm0.070$\\
\hline $\eta \rightarrow \gamma\gamma$  & $-0.265 $  &$ -0.007  $ & $0.015$& $-0.256 $&$ -0.272\pm0.070 $\\
\hline $\eta' \rightarrow \gamma\gamma$ &  $ -0.365 $  &$ 0.011 $
 &     $0.039 $&     $-0.315 $&$ -0.342\pm0.012$\\
\hline
\end{tabular}
\caption{ The predictions of different contributions to the form
factors ${F}_{P \rightarrow \gamma\gamma}$ using the parameter
values in Eq.(\ref{t1k3fit}). All of the values are given in units
of ${\rm GeV}^{-1}$.
 } \label{tableformfactorPgg}
\end{center}
\end{table}

\item  We can predict the decay widths of ${\rho\rightarrow \pi e^{+}e^{-}}$,
${\eta'\rightarrow \gamma e^{+}e^{-}}$ and ${\phi\rightarrow \eta
\mu^{+}\mu^{-}}$ by using the results from the global fit~Eq.\eqref{fit-result}:
\begin{eqnarray}
\Gamma_{\rho\rightarrow \pi e^{+}e^{-}}&=&(3.42 \pm 0.17)\times
10^{-1} {\rm KeV} , \hspace{1.0cm}
 \Gamma_{\eta'\rightarrow \gamma
e^{+}e^{-}}=(8.85 \pm 0.48)\times 10^{-2} {\rm KeV}
\nonumber\\
\Gamma_{\phi\rightarrow \eta \mu^{+}\mu^{-}}&=&(2.22 \pm 0.13)\times
10^{-2} {\rm KeV},
\end{eqnarray}
These predictions are below the upper limits given in PDG~\cite{pdg}.
Hence our results can provide a theoretical hint to the future experimental analysis
on these channels.

\end{enumerate}

\section{ Conclusion }\label{conclude}

In this work, we complete the resonance chiral Lagrangian in the odd
intrinsic parity sector ( VVP type) by including the singlet
$\eta_1$ field as the dynamical degree of freedom. We exploit this
Lagrangian to study radiative decay processes: $P\rightarrow
V\gamma$, $V\rightarrow P\gamma$, $P\rightarrow \gamma\gamma$,
$P\rightarrow \gamma l^{+}l^{-}$, $V\rightarrow Pl^{+}l^{-}$, as
well as the form factors of $\eta\rightarrow \gamma\gamma^{*},
\eta'\rightarrow \gamma\gamma^{*}, \phi\rightarrow
\eta\gamma^{*}$. The two-mixing-angle scheme is used to describe the
$\eta-\eta'$ system in the discussion. By imposing the proper
short distance behavior of QCD, we can fix several combinations of
the unknown resonance couplings. The remaining free resonance
parameters, together with the mixing parameters $F_8, F_0, \theta_8,
\theta_0$ are determined through fitting the various experimental
data. We have shown the resonance chiral Lagrangian can provide a
systematic theoretical framework to handle the various radiative
decay processes involving the resonance states and simultaneously
accommodate the various experimental data. Thus we believe the
$\eta-\eta'$ mixing parameters resulted from this analysis should be
rather reliable.

By integrating out the resonance states in R$\chi$T, we predict the
higher order low energy constants in the odd intrinsic parity
pseudo-Goldstone Lagrangian. We conclude the WZW contribution from
the leading order dominates the processes of $P\to \gamma\gamma$,
with $P=\pi, \eta$ and $\eta'$, by using the $\eta-\eta'$ mixing
parameters from our current analysis. We have also predicted the
decay widths of ${\rho\rightarrow \pi e^{+}e^{-}}$,
${\eta'\rightarrow \gamma e^{+}e^{-}}$ and ${\phi\rightarrow \eta
\mu^{+}\mu^{-}}$, which may shed light on the future measurement for
these three channels.

\section*{Acknowledgments}
We thank Jose Antonio Oller and Juan Jose Sanz-Cillero for discussions. 
This work is supported in part by  National Nature Science
Foundations of China under contract numbers  10925522, 10875001,
11021092 and 11105038. ZHG also acknowledges the grants from Natural Science Foundation of Hebei
Province with contract number A2011205093, Doctor Foundation of Hebei Normal University
with contract number L2010B04, MEC FPA2010-17806 and the Consolider-Ingenio 2010 Programme CPAN (CSD2007-00042).

\appendix

\section{ The decay widths of $ V \to P \gamma$, $ P \to V \gamma$ and $P \to \gamma \gamma $}\label{appendix-vpg-amp}

The various  decay widths from different processes are given below
\begin{eqnarray}
\Gamma(\omega\rightarrow\pi\gamma)&&
=\frac{1}{3}\alpha(\frac{M_\omega^2-m_\pi^2}{2M_\omega})^3 \nonumber
\\&& \bigg\{ -\frac{2\sqrt2 }{F_\pi M_V M_\omega} \big[
(\tilde{c}_1+\tilde{c}_2+8\tilde{c}_3-\tilde{c}_5)m_\pi^2
+(\tilde{c}_2+\tilde{c}_5 -\tilde{c}_1-2\tilde{c}_6)M_\omega^2 \big]
\nonumber \\&& \quad +\frac{4 F_V}{F_\pi M_\omega M_\rho^2} \big[
(\tilde{d}_1+8\tilde{d}_2-\tilde{d}_3)m_\pi^2 +\tilde{d}_3
M_\omega^2  \big] \, \, \bigg\}^2 \,,
\end{eqnarray}
\begin{eqnarray}
\Gamma(\rho^0\rightarrow\pi^0\gamma)&&=
\frac{1}{3}\alpha(\frac{M_\rho^2-m_\pi^2}{2M_\rho})^3 \nonumber \\&&
\bigg\{ -\frac{2\sqrt2 }{3F_\pi M_V M_\rho} \big[
(\tilde{c}_1+\tilde{c}_2+8\tilde{c}_3-\tilde{c}_5)m_\pi^2
+(\tilde{c}_2+\tilde{c}_5 -\tilde{c}_1-2\tilde{c}_6)M_\rho^2 \big]
\nonumber
\\&& \quad +\frac{4 F_V}{3F_\pi M_\rho M_\omega^2} \big[
(\tilde{d}_1+8\tilde{d}_2-\tilde{d}_3)m_\pi^2 +\tilde{d}_3 M_\rho^2
\big] \, \, \bigg\}^2 \,,
\end{eqnarray}
\begin{eqnarray}
\Gamma(K^{*0}\rightarrow K^0\gamma)&&=
\frac{1}{3}\alpha(\frac{M_{K^*}^2-m_K^2}{2M_{K^*}})^3 \nonumber \\&&
\bigg\{ \frac{4\sqrt2 }{3F_\pi M_V M_{K^*}} \big[
(\tilde{c}_1+\tilde{c}_2+8\tilde{c}_3-\tilde{c}_5)m_K^2
+(\tilde{c}_2+\tilde{c}_5 -\tilde{c}_1-2\tilde{c}_6)M_{K^*}^2 \big]
\nonumber \\&& \quad +\frac{2 F_V}{3F_\pi M_{K^*}} \big(
\frac{1}{M_\omega^2} -\frac{3}{M_\rho^2}-\frac{2}{M_\phi^2} \big)
\big[ (\tilde{d}_1+8\tilde{d}_2-\tilde{d}_3)m_K^2 +\tilde{d}_3
M_{K^*}^2 \big] \, \, \bigg\}^2 \,,
\nonumber \\
\end{eqnarray}
\begin{eqnarray}
\Gamma(\rho\rightarrow\eta\gamma)&&=\frac{1}{3}\alpha(\frac{M_\rho^2-m_\eta^2}{2M_\rho})^3
\frac{1}{ \cos{(\theta_0-\theta_8)}^2 }\bigg\{ \nonumber \\ &&
-\frac{2\sqrt2}{\sqrt3 M_V M_\rho}\big(
\frac{\cos\theta_0}{F_8}-\frac{\sqrt2\sin\theta_8}{F_0} \big)\big[
M_\rho^2(\tilde{c}_2-\tilde{c}_1+\tilde{c}_5-2\tilde{c}_6)+m_\eta^2(\tilde{c}_2+\tilde{c}_1-\tilde{c}_5)
+8 \tilde{c}_3 m_\pi^2 \big] \nonumber \\ && + \frac{4 F_V}{\sqrt3
M_\rho^3}\big(
\frac{\cos\theta_0}{F_8}-\frac{\sqrt2\sin\theta_8}{F_0} \big)\big[
\tilde{d}_3(M_\rho^2-m_\eta^2)+\tilde{d}_1m_\eta^2+8 \tilde{d}_2 m_\pi^2 \big] \nonumber \\
&& + \big(-\frac{\sin\theta_8}{F_0} \big) \big[ -\frac{4\sqrt2 M_V}{
M_\rho} \tilde{c}_8+ \frac{8 F_V M_V^2}{ M_\rho^3} \tilde{d}_5 \big]
\bigg\}^2 \,,
\end{eqnarray}
\begin{eqnarray}
\Gamma(\omega\rightarrow\eta\gamma)&&=\frac{1}{9}\Gamma_{\rho\rightarrow\eta\gamma}[M_\rho
\to M_\omega] \nonumber \\&&
=\frac{1}{27}\alpha(\frac{M_\omega^2-m_\eta^2}{2M_\omega})^3
\frac{1}{ \cos{(\theta_0-\theta_8)}^2 }\bigg\{ \nonumber \\ &&
-\frac{2\sqrt2}{\sqrt3 M_V M_\omega}\big(
\frac{\cos\theta_0}{F_8}-\frac{\sqrt2\sin\theta_8}{F_0} \big)\big[
M_\omega^2(\tilde{c}_2-\tilde{c}_1+\tilde{c}_5-2\tilde{c}_6)+m_\eta^2(\tilde{c}_2+\tilde{c}_1-\tilde{c}_5)
+8 \tilde{c}_3 m_\pi^2 \big] \nonumber \\ && + \frac{4 F_V}{\sqrt3
M_\omega^3}\big(
\frac{\cos\theta_0}{F_8}-\frac{\sqrt2\sin\theta_8}{F_0} \big)\big[
\tilde{d}_3(M_\omega^2-m_\eta^2)+\tilde{d}_1m_\eta^2+8 \tilde{d}_2
m_\pi^2 \big] \nonumber
\\ && + \big(-\frac{\sin\theta_8}{F_0} \big) \big[ -\frac{4\sqrt2
M_V}{ M_\omega} \tilde{c}_8+ \frac{8 F_V M_V^2}{ M_\omega^3}
\tilde{d}_5 \big] \bigg\}^2 \,,
\end{eqnarray}
\begin{eqnarray}
\Gamma(\phi\rightarrow\eta\gamma)&&=\frac{1}{3}\alpha(\frac{M_\phi^2-m_\eta^2}{2M_\phi})^3
\frac{1}{ \cos{(\theta_0-\theta_8)}^2 }\bigg\{ \nonumber \\ &&
\frac{4\sqrt2}{3\sqrt3 M_V M_\phi} \times \nonumber \\ && \,\,\times
\big( \frac{\sqrt2\cos\theta_0}{F_8}+\frac{\sin\theta_8}{F_0} \big)
 \big[ M_\phi^2(\tilde{c}_2-\tilde{c}_1+\tilde{c}_5-2\tilde{c}_6)+m_\eta^2(\tilde{c}_2+\tilde{c}_1-\tilde{c}_5) +8 \tilde{c}_3 (2m_K^2-m_\pi^2) \big]
\nonumber \\ && - \frac{8 F_V}{3\sqrt3 M_\phi^3}\big(
\frac{\sqrt2\cos\theta_0}{F_8}+\frac{\sin\theta_8}{F_0} \big) \big[
\tilde{d}_3(M_\phi^2-m_\eta^2)+\tilde{d}_1m_\eta^2+8
\tilde{d}_2(2m_K^2-m_\pi^2) \big] \nonumber \\ && +
\big(-\frac{\sin\theta_8}{F_0} \big) \big[ -\frac{8 M_V}{3 M_\phi}
\tilde{c}_8+ \frac{8\sqrt2 F_V M_V^2}{3 M_\phi^3} \tilde{d}_5 \big]
\bigg\}^2\,,
\end{eqnarray}
\begin{eqnarray}
\Gamma(\phi\rightarrow\eta'\gamma)&&=\frac{1}{3}\alpha(\frac{M_\phi^2-m_{\eta'}^2}{2M_\phi})^3
\frac{1}{ \cos{(\theta_0-\theta_8)}^2 }\bigg\{ \nonumber \\ &&
\frac{4\sqrt2}{3\sqrt3 M_V M_\phi} \times \nonumber \\ && \,\,\times
\big( \frac{\sqrt2\sin\theta_0}{F_8}-\frac{\cos\theta_8}{F_0} \big)
 \big[ M_\phi^2(\tilde{c}_2-\tilde{c}_1+\tilde{c}_5-2\tilde{c}_6)+m_{\eta'}^2(\tilde{c}_2+\tilde{c}_1-\tilde{c}_5) +8 \tilde{c}_3 (2m_K^2-m_\pi^2) \big]
\nonumber \\ && - \frac{8 F_V}{3\sqrt3 M_\phi^3}\big(
\frac{\sqrt2\sin\theta_0}{F_8}-\frac{\cos\theta_8}{F_0} \big) \big[
\tilde{d}_3(M_\phi^2-m_{\eta'}^2)+\tilde{d}_1m_{\eta'}^2+8
\tilde{d}_2(2m_K^2-m_\pi^2) \big] \nonumber \\ && +
\big(\frac{\cos\theta_8}{F_0} \big) \big[ -\frac{8 M_V}{3 M_\phi}
\tilde{c}_8+ \frac{8\sqrt2 F_V M_V^2}{3 M_\phi^3} \tilde{d}_5 \big]
\bigg\}^2\,,
\end{eqnarray}
\begin{eqnarray}
\Gamma(\eta'\rightarrow\omega\gamma)&&=\frac{1}{9}\Gamma_{\eta'\rightarrow\rho\gamma}[M_\rho
\to M_\omega] \nonumber \\&&
=\frac{1}{9}\alpha(\frac{m_{\eta'}^2-M_\omega^2}{2m_{\eta'}})^3
\frac{1}{ \cos{(\theta_0-\theta_8)}^2 }\bigg\{ \nonumber \\ &&
\frac{2\sqrt2}{\sqrt3 M_V M_\omega} \big(
\frac{\sin\theta_0}{F_8}+\frac{\sqrt2\cos\theta_8}{F_0} \big)
 \big[ M_\omega^2(\tilde{c}_2-\tilde{c}_1+\tilde{c}_5-2\tilde{c}_6)+m_{\eta'}^2(\tilde{c}_2+\tilde{c}_1-\tilde{c}_5) +8 \tilde{c}_3 m_\pi^2 \big]
\nonumber \\ && - \frac{4 F_V}{\sqrt3 M_\omega^3} \big(
\frac{\sin\theta_0}{F_8}+\frac{\sqrt2\cos\theta_8}{F_0} \big) \big[
\tilde{d}_3(M_\omega^2-m_{\eta'}^2)+\tilde{d}_1m_{\eta'}^2+8
\tilde{d}_2 m_\pi^2 \big] \nonumber \\ && +
\big(\frac{\cos\theta_8}{F_0} \big) \big[ \frac{4\sqrt2 M_V}{
M_\omega} \tilde{c}_8-\frac{8F_V M_V^2}{ M_\omega^3} \tilde{d}_5
\big] \bigg\}^2\,,
\end{eqnarray}
\begin{eqnarray}
\Gamma(\pi\rightarrow\gamma\gamma)&&= \frac{1}{4}\pi \alpha^2M_\pi^3
\nonumber \\&& \bigg\{-\frac{N_C}{12 \pi^2 F_\pi} -\frac{4\sqrt2
F_V}{3F_\pi M_V} \big( \frac{1}{M_\rho^2} +\frac{1}{M_\omega^2}
\big) (\tilde{c}_1+\tilde{c}_2+8\tilde{c}_3-\tilde{c}_5)m_\pi^2
\nonumber
\\&& \quad +\frac{8 F_V^2}{3F_\pi } \big( \frac{1}{M_\rho^2
M_\omega^2} \big) (\tilde{d}_1+8\tilde{d}_2-\tilde{d}_3)m_\pi^2 \,
\, \bigg\}^2 \,,
\end{eqnarray}
\begin{eqnarray}
\Gamma(\eta\rightarrow\gamma\gamma)&&=  \frac{1}{4}\pi
\alpha^2M_\eta^3 \frac{1}{ \cos{(\theta_0-\theta_8)}^2 } \bigg\{
-\frac{N_C}{12\sqrt3 \pi^2}\big(
\frac{\cos\theta_0}{F_8}-\frac{2\sqrt2\sin\theta_8}{F_0} \big)
\nonumber \\ && -\frac{4\sqrt2\, F_V}{\sqrt3 M_V
}(\frac{1}{M_\rho^2}+\frac{1}{9M_\omega^2})\big[
m_\eta^2(\tilde{c}_2+\tilde{c}_1-\tilde{c}_5) +8 \tilde{c}_3 m_\pi^2
\big] \big( \frac{\cos\theta_0}{F_8}-\frac{\sqrt2\sin\theta_8}{F_0}
\big) \nonumber \\ && +\frac{16\, F_V}{9\sqrt3 M_V
}(\frac{1}{M_\phi^2})\big[
m_\eta^2(\tilde{c}_2+\tilde{c}_1-\tilde{c}_5) +8 \tilde{c}_3
(2m_K^2-m_\pi^2)  \big] \big(
\frac{\sqrt2\cos\theta_0}{F_8}+\frac{\sin\theta_8}{F_0} \big)
\nonumber \\ && + \frac{4 F_V^2}{\sqrt3
}(\frac{1}{M_\rho^4}+\frac{1}{9M_\omega^4})\big[
\tilde{d}_1m_\eta^2-\tilde{d}_3m_\eta^2+8 \tilde{d}_2 m_\pi^2 \big]
\big( \frac{\cos\theta_0}{F_8}-\frac{\sqrt2\sin\theta_8}{F_0} \big)
\nonumber \\ && - \frac{8\sqrt2 F_V^2}{9\sqrt3
}(\frac{1}{M_\phi^4})\big[ \tilde{d}_1m_\eta^2-\tilde{d}_3m_\eta^2+8
\tilde{d}_2(2m_K^2-m_\pi^2) \big] \big(
\frac{\sqrt2\cos\theta_0}{F_8}+\frac{\sin\theta_8}{F_0} \big)
\nonumber \\ && + \big(-\frac{\sin\theta_8}{F_0} \big) \big[
-8\sqrt2 M_V
F_V\tilde{c}_8(\frac{1}{M_\rho^2}+\frac{1}{9M_\omega^2}+\frac{2}{9M_\phi^2})
\nonumber \\ && \qquad\qquad\qquad + 8 F_V^2 M_V^2
\tilde{d}_5(\frac{1}{M_\rho^4}+\frac{1}{9M_\omega^4}+\frac{2}{9M_\phi^4})
\big] \bigg\}^2\,,
\end{eqnarray}
\begin{eqnarray}
\Gamma(\eta'\rightarrow\gamma\gamma)&&=  \frac{1}{4}\pi
\alpha^2M_{\eta'}^3 \frac{1}{ \cos{(\theta_0-\theta_8)}^2 }
\bigg\{ -\frac{N_C}{12\sqrt3 \pi^2}\big(
\frac{\sin\theta_0}{F_8}+\frac{2\sqrt2\cos\theta_8}{F_0} \big)
\nonumber \\ && -\frac{4\sqrt2\, F_V}{\sqrt3 M_V
}(\frac{1}{M_\rho^2}+\frac{1}{9M_\omega^2})\big[
m_{\eta'}^2(\tilde{c}_2+\tilde{c}_1-\tilde{c}_5) +8 \tilde{c}_3
m_\pi^2 \big] \big(
\frac{\sin\theta_0}{F_8}+\frac{\sqrt2\cos\theta_8}{F_0} \big)
\nonumber \\ && +\frac{16\, F_V}{9\sqrt3 M_V
}(\frac{1}{M_\phi^2})\big[
m_{\eta'}^2(\tilde{c}_2+\tilde{c}_1-\tilde{c}_5) +8 \tilde{c}_3
(2m_K^2-m_\pi^2)  \big] \big(
\frac{\sqrt2\sin\theta_0}{F_8}-\frac{\cos\theta_8}{F_0} \big)
\nonumber \\ && + \frac{4 F_V^2}{\sqrt3
}(\frac{1}{M_\rho^4}+\frac{1}{9M_\omega^4})\big[
\tilde{d}_1m_{\eta'}^2-\tilde{d}_3m_{\eta'}^2+8 \tilde{d}_2 m_\pi^2
\big] \big( \frac{\sin\theta_0}{F_8}+\frac{\sqrt2\cos\theta_8}{F_0}
\big) \nonumber \\ && - \frac{8\sqrt2 F_V^2}{9\sqrt3
}(\frac{1}{M_\phi^4})\big[
\tilde{d}_1m_{\eta'}^2-\tilde{d}_3m_{\eta'}^2+8
\tilde{d}_2(2m_K^2-m_\pi^2) \big] \big(
\frac{\sqrt2\sin\theta_0}{F_8}-\frac{\cos\theta_8}{F_0} \big)
\nonumber \\ && + \big(\frac{\cos\theta_8}{F_0} \big) \big[ -8\sqrt2
M_V
F_V\tilde{c}_8(\frac{1}{M_\rho^2}+\frac{1}{9M_\omega^2}+\frac{2}{9M_\phi^2})
\nonumber \\ && \qquad\qquad\qquad + 8 F_V^2 M_V^2
\tilde{d}_5(\frac{1}{M_\rho^4}+\frac{1}{9M_\omega^4}+\frac{2}{9M_\phi^4})
\big] \bigg\}^2\,.
\end{eqnarray}

\section{The form factors of $\phi\rightarrow\eta \gamma^{*}$,
$\eta\rightarrow\gamma \gamma^{*}$ and $\eta'\rightarrow\gamma \gamma^{*}$}\label{appendix-vpgstar-ff}

The definition of the form factor is given in Eq.(\ref{defFF})
and the explicit forms for different processes are reported below
\begin{eqnarray}
F_{\phi\rightarrow\eta \gamma^{*}}(s)&&= \frac{1}{
\cos{(\theta_0-\theta_8)} }\bigg\{ \nonumber \\ &&
\frac{4\sqrt2}{3\sqrt3 M_V M_\phi}\big(
\frac{\sqrt2\cos\theta_0}{F_8}+\frac{\sin\theta_8}{F_0} \big) \times
\nonumber \\ && \,\,\times
 \big[ M_\phi^2(\tilde{c}_2-\tilde{c}_1+\tilde{c}_5-2\tilde{c}_6)+m_\eta^2(\tilde{c}_2+\tilde{c}_1-\tilde{c}_5) +8 \tilde{c}_3 (2m_K^2-m_\pi^2)+ (\tilde{c}_1-\tilde{c}_2+\tilde{c}_5)s  \big]
\nonumber \\ && - \frac{8 F_V}{3\sqrt3 M_\phi} \, D _\phi(s) \,
\big( \frac{\sqrt2\cos\theta_0}{F_8}+\frac{\sin\theta_8}{F_0} \big)
\big[ \tilde{d}_3(M_\phi^2-m_\eta^2 +s )+\tilde{d}_1m_\eta^2+8
\tilde{d}_2(2m_K^2-m_\pi^2) \big] \nonumber \\ && +
\big(-\frac{\sin\theta_8}{F_0} \big) \big[ -\frac{8 M_V}{3 M_\phi}
\tilde{c}_8+ \frac{8\sqrt2 F_V M_V^2}{3 M_\phi} \tilde{d}_5 \, D
_\phi(s) \, \big] \bigg\}\,,
\end{eqnarray}
where the definition of $D_R(s)$ is
\begin{eqnarray}
D_R(s) = \frac{1}{M_R^2 -s - i M_R \Gamma_R(s) } \,.
\end{eqnarray}
For the narrow-width resonances $\omega, \phi$,  we use the constant widths for them
in the numerical discussion. For $\rho$ resonance, the energy dependent width is constructed in the
way introduced in ~\cite{gomez-dumm-00}:
\begin{eqnarray}
\Gamma_{\rho}(s)=\frac{sM_V}{96\pi
F^2}[\sigma_{\pi}^3\theta(s-4m_{\pi}^2)+\frac{1}{2}\sigma_K^3\theta(s-4m_K^2)],
\end{eqnarray}
where $\sigma_P=\sqrt{1-4m_P^2/s}$ and $\theta(s)$ is the step function.

The form factors for $\eta\rightarrow\gamma \gamma^{*}$ and $\eta'\rightarrow\gamma \gamma^{*}$ are
\begin{eqnarray}\label{expetaff}
F_{\eta\rightarrow\gamma \gamma^{*}}(s)&&= \frac{1}{
\cos{(\theta_0-\theta_8)} }\bigg\{ \nonumber \\&&
-\frac{N_C}{12\sqrt3 \pi^2}\big(
\frac{\cos\theta_0}{F_8}-\frac{2\sqrt2\sin\theta_8}{F_0} \big)
\nonumber \\ && -\frac{2\sqrt2\, F_V}{\sqrt3 M_V
}(\frac{1}{M_\rho^2}+\frac{1}{9M_\omega^2})\times \nonumber \\&&
\qquad \big[ m_\eta^2(\tilde{c}_2+\tilde{c}_1-\tilde{c}_5) +8
\tilde{c}_3 m_\pi^2 +(\tilde{c}_1 -\tilde{c}_2 +\tilde{c}_5)s
\,\big] \big(
\frac{\cos\theta_0}{F_8}-\frac{\sqrt2\sin\theta_8}{F_0} \big)
\nonumber \\ && -\frac{2\sqrt2\, F_V}{\sqrt3 M_V }\big[ D_\rho(s) +
\frac{1}{9}D_\omega(s) \big]\times  \nonumber \\&& \qquad \big[
m_\eta^2(\tilde{c}_2+\tilde{c}_1-\tilde{c}_5) +8 \tilde{c}_3 m_\pi^2
+(\tilde{c}_2 +\tilde{c}_5 -\tilde{c}_1-2\tilde{c}_6)s\, \big] \big(
\frac{\cos\theta_0}{F_8}-\frac{\sqrt2\sin\theta_8}{F_0} \big)
\nonumber \\ && +\frac{8\, F_V}{9\sqrt3 M_V
}(\frac{1}{M_\phi^2})\times \nonumber \\&& \quad  \big[
m_\eta^2(\tilde{c}_2+\tilde{c}_1-\tilde{c}_5) +8 \tilde{c}_3
(2m_K^2-m_\pi^2)+(\tilde{c}_1 -\tilde{c}_2 +\tilde{c}_5)s \, \big]
\big( \frac{\sqrt2\cos\theta_0}{F_8}+\frac{\sin\theta_8}{F_0} \big)
\nonumber \\ && +\frac{8\, F_V}{9\sqrt3 M_V }\, D_\phi(s)\, \big(
\frac{\sqrt2\cos\theta_0}{F_8}+\frac{\sin\theta_8}{F_0}
\big)\,\times \nonumber \\&& \qquad  \big[
m_\eta^2(\tilde{c}_2+\tilde{c}_1-\tilde{c}_5) +8 \tilde{c}_3
(2m_K^2-m_\pi^2)+(\tilde{c}_2 +\tilde{c}_5
-\tilde{c}_1-2\tilde{c}_6)s\, \, \big] \nonumber \\ && + \frac{4
F_V^2}{\sqrt3 }\big[ \frac{1}{M_\rho^2} D_\rho(s)
+\frac{1}{9M_\omega^2} D_\omega(s) \big] \big[
\tilde{d}_1m_\eta^2+\tilde{d}_3(s-m_\eta^2)+8 \tilde{d}_2 m_\pi^2
\big] \big( \frac{\cos\theta_0}{F_8}-\frac{\sqrt2\sin\theta_8}{F_0}
\big) \nonumber \\ && - \frac{8\sqrt2 F_V^2}{9\sqrt3 }\big[
\frac{1}{M_\phi^2}\, D_\phi(s) \big] \big[
\tilde{d}_1m_\eta^2+\tilde{d}_3(s-m_\eta^2)+8
\tilde{d}_2(2m_K^2-m_\pi^2) \big] \big(
\frac{\sqrt2\cos\theta_0}{F_8}+\frac{\sin\theta_8}{F_0} \big)
\nonumber \\ && + \big(-\frac{\sin\theta_8}{F_0} \big) \big[
-4\sqrt2 M_V F_V\tilde{c}_8 \big(
D_\rho(s)+\frac{1}{M_\rho^2}+\frac{1}{9}D_\omega(s)+\frac{1}{9M_\omega^2}
+\frac{2}{9}D_\phi(s)+\frac{2}{9M_\phi^2}  \big) \nonumber \\ &&
\qquad\qquad\qquad + 8 F_V^2 M_V^2 \tilde{d}_5 \big(\frac{1}{
M_\rho^2}D_\rho(s)+\frac{1}{9M_\omega^2}D_\omega(s) +D_\phi(s)
\frac{2}{9M_\phi^2}\big) \big] \bigg\}\,,
\end{eqnarray}
\begin{eqnarray}\label{expetapff}
F_{\eta'\rightarrow\gamma \gamma^{*}}(s)&&=\frac{1}{
\cos{(\theta_0-\theta_8)} }\bigg\{ \nonumber \\&&
-\frac{N_C}{12\sqrt3 \pi^2}\big(
\frac{\sin\theta_0}{F_8}+\frac{2\sqrt2\cos\theta_8}{F_0} \big)
\nonumber \\ && -\frac{2\sqrt2\, F_V}{\sqrt3 M_V
}(\frac{1}{M_\rho^2}+\frac{1}{9M_\omega^2})\times \nonumber \\ &&
\qquad   \big[ m_{\eta'}^2(\tilde{c}_2+\tilde{c}_1-\tilde{c}_5)
+8 \tilde{c}_3 m_\pi^2 +(\tilde{c}_1 -\tilde{c}_2 +\tilde{c}_5)s \,
\big] \big( \frac{\sin\theta_0}{F_8}+\frac{\sqrt2\cos\theta_8}{F_0}
\big) \nonumber \\ && -\frac{2\sqrt2\, F_V}{\sqrt3 M_V }\big[
D_\rho(s) + \frac{1}{9}D_\omega(s) \big]\times \nonumber \\ &&
\qquad   \big[ m_{\eta'}^2(\tilde{c}_2+\tilde{c}_1-\tilde{c}_5)
+8 \tilde{c}_3 m_\pi^2 +(\tilde{c}_2 +\tilde{c}_5
-\tilde{c}_1-2\tilde{c}_6)s\, \big] \big(
\frac{\sin\theta_0}{F_8}+\frac{\sqrt2\cos\theta_8}{F_0} \big)
\nonumber \\ && +\frac{8 \, F_V}{9\sqrt3 M_V }
\big(\frac{1}{M_\phi^2}\big) \big(
\frac{\sqrt2\sin\theta_0}{F_8}-\frac{\cos\theta_8}{F_0} \big) \times
\nonumber \\ && \qquad  \big[
m_{\eta'}^2(\tilde{c}_2+\tilde{c}_1-\tilde{c}_5) +8 \tilde{c}_3
(2m_K^2-m_\pi^2) +(\tilde{c}_1 -\tilde{c}_2 +\tilde{c}_5)s \,  \big]
\nonumber \\ && +\frac{8 \, F_V}{9\sqrt3 M_V } \, D_\phi(s) \, \big(
\frac{\sqrt2\sin\theta_0}{F_8}-\frac{\cos\theta_8}{F_0} \big) \times
\nonumber \\ && \qquad   \big[
m_{\eta'}^2(\tilde{c}_2+\tilde{c}_1-\tilde{c}_5) +8 \tilde{c}_3
(2m_K^2-m_\pi^2) +(\tilde{c}_2 +\tilde{c}_5
-\tilde{c}_1-2\tilde{c}_6)s\,\, \big] \nonumber
\\ && + \frac{4 F_V^2}{\sqrt3 }\big[ \frac{1}{M_\rho^2}
D_\rho(s)+\frac{1}{9M_\omega^2} D_\omega(s) \big] \big[
\tilde{d}_1m_{\eta'}^2+\tilde{d}_3(s-m_{\eta'}^2) +8 \tilde{d}_2
m_\pi^2 \big] \big(
\frac{\sin\theta_0}{F_8}+\frac{\sqrt2\cos\theta_8}{F_0} \big)
\nonumber \\ && - \frac{8\sqrt2 F_V^2}{9\sqrt3 }\big[
\frac{1}{M_\phi^2} D_\phi(s) \big] \big[
\tilde{d}_1m_{\eta'}^2+\tilde{d}_3(s-m_{\eta'}^2)+8
\tilde{d}_2(2m_K^2-m_\pi^2) \big] \big(
\frac{\sqrt2\sin\theta_0}{F_8}-\frac{\cos\theta_8}{F_0} \big)
\nonumber \\ && + \big(\frac{\cos\theta_8}{F_0} \big) \big[ -4\sqrt2
M_V F_V\tilde{c}_8 \big(
D_\rho(s)+\frac{1}{M_\rho^2}+\frac{1}{9}D_\omega(s)+\frac{1}{9M_\omega^2}
+\frac{2}{9}D_\phi(s)+\frac{2}{9M_\phi^2}  \big) \nonumber \\ &&
\qquad\qquad\qquad+ 8 F_V^2 M_V^2 \tilde{d}_5 \big(\frac{1}{
M_\rho^2}D_\rho(s)+\frac{1}{9M_\omega^2}D_\omega(s) +D_\phi(s)
\frac{2}{9M_\phi^2}\big) \big] \bigg\}\,.
\end{eqnarray}

\section{ The decay widths of $ V \to P l^- l^+  $, $  P \to V  l^- l^+  $ and $P \to \gamma l^- l^+ $}\label{appendix-vpll}

The kinematic variables used in this sector are defined below.

\begin{itemize}
\item For  $ V (p) \to l^- (k_1) \, l^+ (k_2) \, P (q)$
\begin{eqnarray}
k&=&k_1+k_2 \,, \nonumber \\
(k_1+k_2)^2&=&s  \,, \nonumber \\
(q+k_2)^2&=&t  \,. \nonumber \\
\end{eqnarray}
The general decay amplitude is
\begin{eqnarray}
T( V \to l^- l^+ P )= e^2
\epsilon_{\mu\nu\rho\sigma}\epsilon_p^{\nu}k^{\rho} p^{\sigma}\,
\bar{u}(k_1) \gamma^\mu v(k_2)\, A_{ V \to l^- l^+ P}(s,t).
\end{eqnarray}
To calculate the decay width, one needs
\begin{eqnarray}
&&|\epsilon_{\mu\nu\rho\sigma}\epsilon_p^{\nu}k^{\rho} p^{\sigma}\,
\bar{u}(k_1) \gamma^\mu v(k_2)|^2 = \nonumber \\&& 2 m_l^4 s +
 2 m_l^2 \big[ m_P^4 + M_V^4 - M_V^2 s - m_P^2 (2 M_V^2 + s) - 2 s \,t \big]
\nonumber \\&& + s \big[ m_P^4 + M_V^4 + s^2 + 2 s\, t + 2 t^2 - 2
m_P^2 (s + t) - 2 M_V^2 (s + t) \big]\,.
\end{eqnarray}

Then the decay width is found to be
\begin{eqnarray}
\Gamma_{ V \to P l^- l^+}= \frac{1}{3} \frac{1}{8\pi^3}\frac{1}{32\,
M_V} \int^{(M_V-m_P)^2}_{4m_l^2} d s  \int^{\rm t_{max}}_{\rm
t_{min}} d t \, |T(s,t)|^2
\end{eqnarray}
where
\begin{eqnarray}
{\rm t_{min}} &=& \frac{M_V^2 +m_P^2 +2m_l^2 -s}{2} -
\frac{\sqrt{s(s-4m_l^2)[s-(M_V+m_P)^2][s-(M_V-m_P)^2]}}{2s} \,,
\nonumber \\
{\rm t_{max}} &=& \frac{M_V^2 +m_P^2 +2m_l^2 -s}{2} +
\frac{\sqrt{s(s-4m_l^2)[s-(M_V+m_P)^2][s-(M_V-m_P)^2]}}{2s} \,.
\nonumber \\
\end{eqnarray}

\item For  $  P (q) \to l^- (k_1) \, l^+ (k_2) V (p) \, $
\begin{eqnarray}
k&=&k_1+k_2 \,, \nonumber \\
(k_1+k_2)^2&=&s  \,, \nonumber \\
(p+k_2)^2&=&t  \,, \nonumber \\
\end{eqnarray}

The general decay amplitude is
\begin{eqnarray}
T( P \to l^- l^+ V )= e^2
\epsilon_{\mu\nu\rho\sigma}\epsilon_p^{\nu}k^{\rho} p^{\sigma}\,
\bar{u}(k_1) \gamma^\mu v(k_2)\, A_{ P \to l^- l^+ V }(s,t).
\end{eqnarray}
To calculate the decay width, one needs
\begin{eqnarray}
&&|\epsilon_{\mu\nu\rho\sigma}\epsilon_p^{\nu}k^{\rho} p^{\sigma}\,
\bar{u}(k_1) \gamma^\mu v(k_2)|^2 = \nonumber \\&& 2 m_l^4 s +
 2 m_l^2 \big[ m_P^4 + M_V^4 - M_V^2 s - m_P^2 (2 M_V^2 + s) - 2 s \,t \big]
\nonumber \\&& + s \big[ m_P^4 + M_V^4 + s^2 + 2 s\, t + 2 t^2 - 2
m_P^2 (s + t) - 2 M_V^2 (s + t) \big]\,.
\end{eqnarray}

The decay width is
\begin{eqnarray}
\Gamma_{ P \to V l^- l^+}= \frac{1}{8\pi^3}\frac{1}{32\, m_P}
\int^{(m_P-M_V)^2}_{4m_l^2} d s  \int^{\rm t_{max}}_{\rm t_{min}} d
t \, |T(s,t)|^2
\end{eqnarray}
where
\begin{eqnarray}
{\rm t_{min}} &=& \frac{M_V^2 +m_P^2 +2m_l^2 -s}{2} -
\frac{\sqrt{s(s-4m_l^2)[s-(M_V+m_P)^2][s-(M_V-m_P)^2]}}{2s} \,,
\nonumber \\
{\rm t_{max}} &=& \frac{M_V^2 +m_P^2 +2m_l^2 -s}{2} +
\frac{\sqrt{s(s-4m_l^2)[s-(M_V+m_P)^2][s-(M_V-m_P)^2]}}{2s} \,.
\nonumber \\
\end{eqnarray}

\item For  $  P (p) \to l^- (k_{21}) \, l^+ (k_{22}) \gamma (k_1) \, $
\begin{eqnarray}
k_2&=&k_{21}+k_{22} \,, \nonumber \\
(k_{21}+k_{22})^2&=&s  \,, \nonumber \\
(k_1+k_{22})^2&=&t  \,. \nonumber \\
\end{eqnarray}

The general decay amplitude is
\begin{eqnarray}
T( P \to l^- l^+ \gamma )= e^3
\epsilon_{\mu\nu\rho\sigma}\epsilon_1^{\nu}k_1^{\rho}k_2^{\sigma}\,
\bar{u}(k_{21}) \gamma^\mu v(k_{22})\, A_{ P \to l^- l^+
\gamma}(s,t).
\end{eqnarray}

To calculate the decay width, one needs
\begin{eqnarray}
&&|\epsilon_{\mu\nu\rho\sigma}\epsilon_1^{\nu}k_1^{\rho}k_2^{\sigma}\,
\bar{u}(k_{21}) \gamma^\mu v(k_{22})|^2 = \nonumber \\&& 2 m_l^4 s +
2 m_l^2 ( m_P^4 - m_P^2 s - 2 s \,t) +
  s [m_P^4 + s^2 + 2 s\, t + 2 t^2 - 2 m_P^2 (s + t)]  \,. \nonumber \\
\end{eqnarray}

The decay width is
\begin{eqnarray}
\Gamma_{ P \to \gamma l^- l^+}= \frac{1}{8\pi^3}\frac{1}{32\, m_P}
\int^{m_P^2}_{4m_l^2} d s  \int^{\rm tmax}_{\rm tmin} d t \,
|T(s,t)|^2
\end{eqnarray}
where
\begin{eqnarray}
{\rm tmin} &=& \frac{m_P^2 +2m_l^2 -s}{2} -
\frac{(m_P^2-s)\sqrt{s(s-4m_l^2)}}{2s} \,,
\nonumber \\
{\rm tmax} &=& \frac{m_P^2 +2m_l^2 -s}{2} +
\frac{(m_P^2-s)\sqrt{s(s-4m_l^2)}}{2s} \,.
\nonumber \\
\end{eqnarray}

\end{itemize}

The explicit expressions of the decay amplitudes are given below
\begin{eqnarray}
T(\omega \rightarrow\pi l^- l^+)&&= -e^2
\epsilon_{\mu\nu\rho\sigma}\epsilon_p^{\nu}k^{\rho} p^{\sigma}\,
\bar{u}(k_1) \gamma^\mu v(k_2)\, \frac{1}{s}\,\bigg\{ \nonumber \\&&
-\frac{2\sqrt2 }{F_\pi M_V M_\omega} \big[
(\tilde{c}_1+\tilde{c}_2+8\tilde{c}_3-\tilde{c}_5)m_\pi^2
+(\tilde{c}_2+\tilde{c}_5 -\tilde{c}_1-2\tilde{c}_6)M_\omega^2
+(\tilde{c}_1-\tilde{c}_2+\tilde{c}_5)s \,\big] \nonumber
\\&& +\frac{4 F_V}{F_\pi M_\omega }\, D_\rho(s) \, \big[
(\tilde{d}_1+8\tilde{d}_2-\tilde{d}_3)m_\pi^2 +\tilde{d}_3
(M_\omega^2 + s) \big]\, \bigg\}\,,
\end{eqnarray}
\begin{eqnarray}
T(\rho \rightarrow\pi l^- l^+)&&= -e^2
\epsilon_{\mu\nu\rho\sigma}\epsilon_p^{\nu}k^{\rho} p^{\sigma}\,
\bar{u}(k_1) \gamma^\mu v(k_2)\, \frac{1}{s}\,\bigg\{ \nonumber \\&&
-\frac{2\sqrt2 }{3F_\pi M_V M_\rho} \big[
(\tilde{c}_1+\tilde{c}_2+8\tilde{c}_3-\tilde{c}_5)m_\pi^2
+(\tilde{c}_2+\tilde{c}_5 -\tilde{c}_1-2\tilde{c}_6)M_\rho^2
+(\tilde{c}_1-\tilde{c}_2+\tilde{c}_5)s \,\big] \nonumber
\\&& +\frac{4 F_V}{3F_\pi M_\rho }\, D_\omega(s) \, \big[
(\tilde{d}_1+8\tilde{d}_2-\tilde{d}_3)m_\pi^2 +\tilde{d}_3 (M_\rho^2
+ s) \big]\, \bigg\}\,,
\end{eqnarray}
\begin{eqnarray}
T(\phi\rightarrow\eta l^- l^+)&&= -e^2
\epsilon_{\mu\nu\rho\sigma}\epsilon_p^{\nu}k^{\rho} p^{\sigma}\,
\bar{u}(k_1) \gamma^\mu v(k_2)\, \frac{1}{s} \,
F_{\phi\rightarrow\eta \gamma^{*}}(s),
\end{eqnarray}
\begin{eqnarray}
T(\pi\rightarrow\gamma l^- l^+)&&= e^3
\epsilon_{\mu\nu\rho\sigma}\epsilon_1^{\nu}k_1^{\rho}k_2^{\sigma}\,
\bar{u}(k_{21}) \gamma^\mu v(k_{22})\, \frac{1}{s} \,\bigg\{
\nonumber \\&& -\frac{N_C}{12 \pi^2 F_\pi} -\frac{2\sqrt2
F_V}{3F_\pi M_V} \big( \frac{1}{M_\rho^2} +\frac{1}{M_\omega^2}
\big) \big[
(\tilde{c}_1+\tilde{c}_2+8\tilde{c}_3-\tilde{c}_5)m_\pi^2
+(\tilde{c}_1 -\tilde{c}_2 +\tilde{c}_5)s \big] \nonumber \\&&
-\frac{2\sqrt2 F_V}{3F_\pi M_V} \big[ D_\rho(s) + D_\omega(s) \big]
\big[ (\tilde{c}_1+\tilde{c}_2+8\tilde{c}_3-\tilde{c}_5)m_\pi^2
+(\tilde{c}_2 +\tilde{c}_5 -\tilde{c}_1-2\tilde{c}_6)s \big]
\nonumber \\&& +\frac{4 F_V^2}{3F_\pi } \big[ \frac{1}{M_\rho^2}\,
D_\omega(s) +\frac{1}{M_\omega^2}\, D_\rho(s) \big] \big[
(\tilde{d}_1+8\tilde{d}_2-\tilde{d}_3)m_\pi^2 +\tilde{d}_3 s \big]
\, \, \bigg\}
\end{eqnarray}
\begin{eqnarray}
T(\eta\rightarrow\gamma  l^- l^+)&&= e^3
\epsilon_{\mu\nu\rho\sigma}\epsilon_1^{\nu}k_1^{\rho}k_2^{\sigma}\,
\bar{u}(k_{21}) \gamma^\mu v(k_{22})\,
\frac{1}{s}F_{\eta\rightarrow\gamma \gamma^{*}}(s),
\end{eqnarray}
\begin{eqnarray}
T(\eta'\rightarrow\gamma l^- l^+)&&= e^3
\epsilon_{\mu\nu\rho\sigma}\epsilon_1^{\nu}k_1^{\rho}k_2^{\sigma}\,
\bar{u}(k_{21}) \gamma^\mu v(k_{22})\,
\frac{1}{s}F_{\eta'\rightarrow\gamma \gamma^{*}}(s).
\end{eqnarray}
where $F_{\eta\rightarrow\gamma \gamma^{*}}$
and $F_{\eta'\rightarrow\gamma \gamma^{*}}$ are given in
Eqs.\eqref{expetaff} and \eqref{expetapff}.

%\end{CJK}
\end{document}